\documentclass[12pt]{article}
\usepackage{epsf,amsfonts,amssymb,epsfig,amsmath,mathtools,graphics,slashed}
\usepackage{hyperref,hep,graphicx,subfig}
\usepackage{rotating}
\usepackage[final]{showlabels}
\usepackage{xcolor}
\allowdisplaybreaks

\definecolor{greenish}{RGB}{0,150,0}

\definecolor{yellowish}{RGB}{150,150,0}

\newcommand{\nn}{\notag \\}

\makeatletter

\@addtoreset{equation}{section}
\makeatother

\begin{document}

\begin{titlepage}

\vfill

\begin{flushright}
DCPT-19/15
\end{flushright}

\vfill

\begin{center}
   \baselineskip=16pt
   {\Large\bf Incoherent hydrodynamics and density waves}
  \vskip 1.5cm
  \vskip 1.5cm
      Aristomenis Donos$^1$, Daniel Martin$^1$, Christiana Pantelidou$^1$ and Vaios Ziogas$^2$\\  
    \vskip .6cm
      \begin{small}
      \textit{$^1$ Centre for Particle Theory and Department of Mathematical Sciences,\\ Durham University,
       Durham, DH1 3LE, U.K.}
        \end{small}\\
      \vskip .6cm
      \begin{small}
      \textit{$^2$Shanghai Center for Complex Physics, School of Physics and Astronomy,\\
        Shanghai Jiao Tong University, Shanghai 200240, China}
        \end{small}\\
         
\end{center}

\vfill

\begin{center}
\textbf{Abstract}
\end{center}
\begin{quote}
We consider thermal phases of holographic lattices at finite chemical potential in which a continuous internal bulk symmetry can be spontaneously broken. In the normal phase, translational symmetry is explicitly broken by the lattice and the only conserved quantities are related to time translations and the electric charge. The long wavelength excitations of the corresponding charge densities are described by incoherent hydrodynamics yielding two perturbative modes which are diffusive. In the broken phase an additional hydrodynamic degree of freedom couples to the local chemical potential and temperature and we write an effective theory describing the coupled system at leading order in a derivative expansion.
\end{quote}

\vfill

\end{titlepage}

\setcounter{equation}{0}

\section{Introduction}

The AdS/CFT correspondence provides a consistent framework to study universal features of large classes of strongly coupled field theories with a holographic dual. In the limit of classical gravity in particular, we can carry out straightforward computations in the bulk which are of high physical significance for the conformal field theory. Strong coupling is a fundamental difficulty arising in the theoretical understanding of certain classes of condensed matter systems such as the cuprate superconductors. The reduced amount of symmetry in those systems makes the field theoretical approach even less constrained. From this perspective, holography is an invaluable source of information as the duality makes possible the study of RG flows of strongly coupled theories with little or no symmetry through their gravity duals.

Having in mind applications in condensed matter physics \cite{Herzog:2009xv,Hartnoll:2009sz,Hartnoll:2016apf}, we will consider classes of strongly coupled theories at finite chemical potential and temperature. In order to eliminate momentum from the conserved charges of the system, spatial translations will be explicitly broken by a holographic lattice. This is necessary for the heat current to relax and the zero frequency transport coefficients to be finite. In order to make the analysis more tractable, we will implement the lattice via a Q-lattice construction which requires the presence of global $U(1)$ symmetries in the bulk \cite{Donos:2013eha} . %When the strength of the holographic lattice is taken to be parametrically small, momentum is almost conserved and the system will have a well defined rate of momentum relaxation. However, this is not going to be necessary for our construction.

At low temperatures, the systems we consider can spontaneously break these global symmetries in the bulk, giving rise to additional hydrodynamic degrees of freedom \cite{Donos:2019tmo,Donos:2019txg}. This setup will allow us to model holographic transport which captures the effects of the coupling between the currents and the sliding density wave. Moreover, we can add a small source which breaks the internal symmetry, gapping the sliding mode of the density wave in a controlled manner. Such a scenario has been considered before \cite{Andrade:2017ghg,Andrade:2017leb,Andrade:2017cnc,Andrade:2015iyf,Amoretti:2018tzw,Andrade:2018gqk} , the advantage of our setup is that we can have control over the separation between the momentum and the phase relaxation times. The effect of this small pinning parameter on the finite frequency transport properties is of interest in condensed matter physics because of the transfer of spectral weight to energy scales set by the gap. Such effects are expected to play a prominent role in systems that include the pseudogap region of the hight-$T_c$ phase diagram and bad metals \cite{vojta,Gruner:1988zz,Gruner:1994zz}.

One of the key concepts in condensed matter systems is the dynamics of excitations at wavelengths much bigger than any other scale of the system. In general, such excitations are captured by hydrodynamics which provides an effective description in a derivative expansion. In that regime, one can argue that late time dynamics is governed by conserved charges and potential gapless modes emerging from broken symmetries. Understanding the dominant mechanisms in such processes is of physical significance as they will also determine the low frequency transport properties when the system is driven by external sources.

At temperatures higher than then critical one, hydrodynamics is dictated by the only conserved quantities of the system which are related to time translations and the electric charge. In this regime, the long wavelength excitations are effectively described by incoherent hydrodynamics \cite{Hartnoll:2014lpa,Donos:2017gej}. A good set of local variables which capture the dynamics of these excitations are the local temperature and chemical potential. At the level of linear response, the system is then dominated by two thermoelectric modes which are diffusive. Note in particular that relativistic systems with weak momentum relaxation can be described by using relativistic hydrodynamics with perturbative deformations which break translations \cite{Davison:2014lua,Donos:2017gej}. In that context, the local fluid velocity can be integrated out as a result of momentum relaxation; this was discussed in \cite{Donos:2017gej} from the field theory perspective and in \cite{Donos:2017ihe} within holography. The advantage of the class of models we will consider in our paper is that the momentum relaxation mechanisms will not have to be perturbatively small. Nevertheless, we will still be able to capture the physics of depinning of density waves.

In the broken phase, incoherent hydrodynamics needs to be supplemented by an additional variable which captures the dynamics of the emergent gapless mode due to symmetry breaking in the bulk. As shown in \cite{Donos:2019txg}, when the order parameter doesn't itself break translations, this mode decouples from the other gapless modes of the system and is diffusive. However, in the system we will consider in this paper, this novel mode will couple to the heat and electric currents and for this reason we will call it ``sliding'' in the present context. One therefore anticipates that below the critical temperature two thermoelectric modes and the extra mode due to symmetry breaking will mix in to yield three diffusive ones.

By carefully examining our holographic system through techniques very similar to those of \cite{Donos:2017ihe}, we will manage to extract the dispersion relations of the three anticipated modes\footnote{Here we will consider only longitudinal excitations with wavevectors parallel to the thermal and electric currents. Transverse modes can be studied by using very similar techniques and we will leave that for future work.}. However, here we will technically approach the same problem in two different but equivalent ways. In both cases we will use the bulk solutions generated by varying the thermodynamic backgrounds with respect to the global temperature, chemical potential and phase of the broken bulk $U(1)$ as seed solutions to build our derivative expansion. As one might expect for a relativistic system, in building the hydrodynamic description we will encounter a vector variable, which can be seen as a fluid velocity on the event horizon and three scalars which are the local temperature, chemical potential and phase of the VEV of a complex scalar which has condensed. However, with translations being broken in our system, we will manage to integrate out the vector variable from the description ending up with the three scalars that will be our variables for the incoherent hydrodynamics. As we will see in the main text, by integrating one of the radial equations, we will obtain a Josephson-type equation for the phase of the complex scalar. We will therefore need to identify two additional scalar equations that will fully determine the time evolution of the local temperature and chemical potential.

From the classical gravity point of view, the most effective and natural way to obtain a closed set of equations for the three scalar variables is to impose the diffeomorphism and Gauss constraints close to the horizon of the black hole \cite{Donos:2015gia,Banks:2015wha,Donos:2017ihe}. This will be our first approach. An equivalent but field theoretically more telling way to view the same constraints is to impose them close to the boundary of the spacetime. In this limit, they have the interpretation of the Ward identities of the stress tensor and the global $U(1)$ charge conservation. To do this, we will give the constitutive relations for the electric and heat currents in terms of gradients of our three scalar hydrodynamic variables and a number of transport coefficients which will be determined by the black hole horizon.

Moreover, we will introduce finite frequency boundary sources that correspond to temperature gradient and electric field that will also enter our hydrodynamic description. Even more interestingly, we will include a perturbative source on the boundary which pins down the sliding density wave and which will also appear in our theory \cite{Donos:2019tmo,Donos:2019txg}. Using our results we will obtain an analytic formula for the transport coefficients that were numerically computed in \cite{Donos:2019tmo} up to frequencies set by the pinning scale.

Finally, we perform non-trivial numerical checks of our analytic results for a specific Q-lattice model which realises the breaking of a global bulk symmetry. More specifically, for small wavelengths we will numerically identify the three diffusive modes we anticipate from our analysis. This will allow us to extract the corresponding diffusion constants and match them with our analytic expressions. As a further check, for relatively weak lattices we will confirm that as we take wavelengths short enough to be comparable to the momentum relaxation time scale, one of our diffusive poles collides with the momentum relaxation pole to produce two sound modes \cite{Davison:2014lua,Grozdanov:2018fic}. Finally, we confirm our results for the gap and the low-frequency AC thermoelectric conductivities by comparing with  the numerical data of \cite{Donos:2019tmo}.

Our paper is organised as follows. In section \ref{sec:setup} we present the class of holographic models that capture the physics we are interested in and we give some details about the phase transition and thermodynamics. In section \ref{sec:pert} we take the system to be in its broken phase and we study the three hydrodynamic modes we described earlier in the introduction. An essential element of our analysis will be the infinitely long wavelength solutions which we generate by variations with respect to the thermodynamic variables of the backgrounds we describe in section \ref{sec:setup}. The analysis in this section is based on imposing the gravitational constraints on the black hole horizon. Section \ref{sec:hydro} is devoted to deriving the constitutive relations for the currents of the boundary theory and introducing appropriate sources. This will help us develop to the enlarged version of incoherent hydrodynamics describing our system. In section \ref{sec:pinning} we will introduce a pinning parameter for the density wave and we will compute the frequency dependent retarded Green's functions for our system by using standard techniques. This will allow us to quantitatively explain the transfer of spectral weight due to the gapped mode in our system and explain the optical conductivities that were computed numerically in \cite{Donos:2019tmo}. Finally, in section \ref{sec:numerics} we will perform a number of numerical checks for the analytic formulae for the diffusion constants we will derive in section \ref{sec:pert} and the gap. We conclude in section \ref{sec:discussion} with a discussion.

\section{Set-up}\label{sec:setup}

In this section we will discuss a class of four dimensional holographic models in which we can introduce a chemical potential, momentum relaxation and global bulk symmetries in the simplest way possible.

We wish to use 2 complex scalars $Y_{I}$, $I=1,2$, dual to either marginal or relevant operators, as a Q-lattice to explicitly break translations \cite{Donos:2013eha}. As we will later see, each one of those complex scalar will be used to break translations in each of the spatial direction, $x_1,x_2$. Moreover, we want to realise the spontaneous breaking of global $U(1)$'s in the bulk, for which we use 2 complex scalars $Z_{I}$, $I=1,2$. The only restriction on the conformal dimensions of the field theory duals of $Z_{I}$ that we need will come when we consider the effects of pinning later in our paper where we will need to introduce a perturbative static source for $Z_{I}$. Similarly to the $Y_{I}$'s, each one of $Z_{I}$'s will spontaneously break translations in a spatial different direction.

The bulk action which captures all the necessary ingredients is
\begin{align}\label{eq:bulk_action}
S_{bulk}&=\int d^4 x \sqrt{-g}\,\Bigl(R-V-\frac{1}{2}\sum_{I}\left(G_{I}\,\partial Z_{I}\partial \bar{Z}_{I} +W_{I}\,\partial Y_{I}\partial \bar{Y}_{I} \right) -\frac{\tau}{4}\,F^{2} \Bigr) \,,
\end{align}
If we demand that the functions $V$, $G_{I}$, $W_{I}$ and $\tau$ only depend on the squares of the moduli $b_{I}=Z_{I}\bar{Z}_{I}$ and $n_{I}=Y_{I}\bar{Y}_{I}$ of the complex scalars, our theory will realise the four global $U(1)$'s that we will need for our construction. Under this restriction, the variation of \eqref{eq:bulk_action} yields the equations of motion
\begin{align}\label{eq:eom1}
&L_{\mu\nu}\equiv R_{\mu\nu}-\frac{\tau}{2} (F_{\mu\rho}F_{\nu}{}^{\rho}-\frac{1}{4}g_{\mu\nu}F^2)-\frac{1}{2}g_{\mu\nu} V-\frac{1}{2}\sum_{I}\left( G_{I}\, \partial_{(\mu}Z_{I}\partial_{\nu)}\bar{Z}_{I}+ W_{I}\, \partial_{(\mu}Y_{I}\partial_{\nu)}\bar{Y}_{I}\right)=0\,,\notag\\
&\nabla^{\mu}\left(G_{J} \nabla_{\mu}Z_{J}\right)-\partial_{b_{J}}V\,Z_{J}-\sum_{I}\left(\partial_{b_{J}}G_{I}\,\partial Z_{I}\partial \bar{Z}_{I} +\partial_{b_{J}}W_{I}\,\partial Y_{I}\partial \bar{Y}_{I} \right)\,Z_{J}-\frac{\partial_{b_{J}}\tau}{4}Z_{J}\,F^{2}=0\,\notag\,,\\
&\nabla^{\mu}\left(W_{J} \nabla_{\mu}Y_{J}\right)-\partial_{n_{J}}V\,Y_{J}-\sum_{I}\left(\partial_{n_{J}}G_{I}\,\partial Z_{I}\partial \bar{Z}_{I} +\partial_{n_{J}}W_{I}\,\partial Y_{I}\partial \bar{Y}_{I} \right)\,Y_{J}-\frac{\partial_{n_{J}}\tau}{4}Y_{J}\,F^{2}=0\,\notag\,,\\
&C^{\nu}\equiv \nabla_{\mu}\left( \tau\,F^{\mu\nu}\right) =0\,.
\end{align}
Moreover, by requiring that for small values of the scalars the functions appearing in our action \eqref{eq:bulk_action} behave as,
\begin{align}
V=-6+\frac{1}{2}\sum_{I}\left(m_{Z_{I}}^{2}\,Z_{I}\bar{Z}_{I}+m_{Y_{I}}^{2}\,Y_{I}\bar{Y}_{I} \right)+\cdots\notag\\
G_{I}=1+\cdots,\quad W_{I}=1+\cdots,\quad \tau=1+\cdots\,,
\end{align}
the equations of motion \eqref{eq:eom1} admit as a solution the unit radius $AdS_{4}$ metric
\begin{align}\label{eq:ads}
ds_{4}^{2}=r^{2}(-dt^{2}+dx_{1}^{2}+dx_{2}^{2})+\frac{dr^{2}}{r^{2}}\,,\quad Z_{I}=Y_{I}=A_{\mu}=0\,.
\end{align}
Given the above restrictions, the field theory duals of $Z_{I}$ and $Y_{I}$ will correspond to operators of dimensions $\Delta_{Z_{I}}$ and $\Delta_{W_{I}}$ with $\Delta_{Z_{I}}\,(\Delta_{Z_{I}}-3)=m_{Z_{I}}^{2}$ and $\Delta_{Y_{I}}\,(\Delta_{Y_{I}}-3)=m_{Y_{I}}^{2}$.

In the rest of the paper we will find it convenient to parametrise the scalars in polar coordinates according to\footnote{For our purposes, we will not need the full non-linear transformation \eqref{eq:polar_decomposition}. An equivalent way to derive all the results in our paper is to consider perturbations for the complex scalar fields of the form $\delta Z=\phi_b e^{i\chi_b}\,i\,\delta \chi+e^{i\chi_b}\,\delta \phi\,$ around backgrounds with $Z_{b}=e^{i\chi_{b}}\phi_{b}$.}
\begin{align}\label{eq:polar_decomposition}
Y_{I}=\psi_{I}\,e^{i \sigma_{I}},\quad Z_{I}=\phi_{I}\,e^{i \chi_{I}}\,,
\end{align}
bringing the action \eqref{eq:bulk_action} to the form
\begin{align}\label{eq:bulk_action_alt}
S_{bulk}&=\int d^4 x \sqrt{-g}\,\Bigl(R-V-\frac{1}{2}\sum_{I}\left(G_{I}\,(\partial \psi_{I})^{2} +W_{I}\,(\partial \phi_{I})^{2} \right)\notag\\
&\qquad\qquad -\frac{1}{2}\sum_{I}\left(\Psi_{I}\,(\partial \sigma_{I})^{2} +\Phi_{I}\,(\partial \chi_{I})^{2} \right)-\frac{\tau}{4}\,F^{2} \Bigr)\,,\notag\\
\Psi_{I}&\equiv G_{I}\,\psi_{I}{}^{2},\quad \Phi_{I}\equiv W_{I}\,\phi_{I}{}^{2}\,.
\end{align}
Notice that in this parametrisation the functions $V$, $\tau$, $G_{I}$, $W_{I}$, and therefore $\Psi_{I}$ and $\Phi_{I}$ do not depend on $\sigma_{I}$ and $\chi_{I}$. The global $U(1)$'s in the bulk are captured by the shift symmetries of $\sigma_{I}$ and $\chi_{I}$ and by the fact that we should make the identifications $\sigma_{I} \sim \sigma_{I}+2\pi$ and $\chi_{I} \sim \chi_{I}+2\pi$ for the target space of the sigma model to be regular. The equations of motion coming from \eqref{eq:bulk_action_alt} are equivalent to those coming from \eqref{eq:bulk_action}; for later reference we write here the ones coming from a variation with respect to $\chi_{I}$,
\begin{align}\label{eq:bulk_eom_chi}
\nabla^{\mu}\left( \Phi_{I}\nabla_{\mu}\chi_{I}\right)=&0\,.
\end{align}

Introducing a chemical potential and lattice deformation to our boundary theory will modify the background bulk metric from that of $AdS_{4}$ in \eqref{eq:ads}. However, with our Q-lattice construction we will be able to maintain homogeneity and therefore avoid the problem of having to solve PDEs. As we will consider thermal states,  finite temperature will require the existence of an event horizon which we assume to be of planar topology. These black holes will describe the normal phase of our system. In addition, we will consider a density wave state that will appear spontaneously for $T<T_c$ and will also implement it through a Q-lattice construction. An ansatz which captures all the necessary ingredients, including the spontaneous breaking of the $U(1)$ related to shifts of $\chi_{I}$ is given by
\begin{align} \label{eq:DC_ansatz}
ds^{2}&=-U(r)\,dt^{2}+\frac{1}{U(r)}\,dr^{2}+e^{2V_{1}(r)}\,dx^1 dx^1 +e^{2V_{2}(r)}\,dx^2 dx^2\,,\nn
A&=a(r)\,dt\,,\nn
\phi_{I}&=\phi_{I}(r)\,,\qquad\qquad \chi_{I}=k_{Ii} x^{i}+c_{I}\,,\nn
\psi_{I}&=\psi_{I}(r)\,,\qquad\qquad \sigma_{I}=k_{sIi} x^{i}\,,
\end{align}
where we shall take the case $k_{Ii}=k_{i}\,\delta^{i}_{I}$ and $k_{sIi}=k_{si}\,\delta^{i}_{I}$ (no summation). This particular choice of wavevectors associates each of the four complex scalars to a spatial direction. As we can see, the constants $c_{I}$ that we can freely choose in \eqref{eq:DC_ansatz} represent the Goldstone modes in the bulk due to symmetry breaking and cannot be fixed by boundary conditions. Since these modes shift the density wave, they can be interpreted as sliding modes. In order to introduce the appropriate sources and accommodate the right VEVs we demand the near conformal boundary expansion 
\begin{align}\label{asymptsol}
U&\to (r+R)^2+\cdots+W\,(r+R)^{-1}+\cdots,\qquad V_{1}\to \log(r+R)+\cdots+W_p (r+R)^{-3}+\cdots,\nn V_{2}&\to \log(r+R)+\cdots,\qquad\qquad\qquad a\to\mu+Q\,(r+R)^{-1}+\cdots,\nn
\psi_{I}&\to  \psi_{Is}\,(r+R)^{-3+\Delta_{Y_{I}}}+\cdots+\psi_{Iv}\,(r+R)^{-\Delta_{Y_{I}}}+\cdots, \qquad \phi_{I}\to  \phi_{Iv}\,(r+R)^{-\Delta_{Z_{I}}}+\cdots\,.
\end{align}
The constant of integration $R$ that appears in the above expansion represents the part of reparametrisation invariance which is left unfixed by the ansatz \eqref{eq:DC_ansatz}, given by constant shifts of the radial coordinate. We will choose $R$ so that the horizon of the black hole is located at $r=0$. The expansion \eqref{asymptsol} reflects the chemical potential $\mu$ while the constants $\psi_{Is}$ represent the strength of the explicit breaking of translations due to the Q-lattice and they are all meant to be fixed as deformation parameters of the theory. Moreover, the internal $U(1)$s associated with $Z_{I}$ are spontaneously broken whenever our solutions have $ \phi_{Iv}\neq 0$ which we  expect to happen generically at temperatures below a critical one. It is useful to note that the VEVs $\langle \mathcal{O}_{Z_{I}}\rangle$ of the duals of $Z_{I}$ are $\langle \mathcal{O}_{Z_{I}}\rangle=(\Delta_{Z_{I}}-3/2)\,\phi_{Iv}\,e^{ik_{i}x^{i}+i\,c_{I}}$ implying that $k_{i}$ are not to be fixed by hand. Most importantly, for the background with $k_{i}\neq 0$, the order parameter which breaks the internal $U(1)$'s also breaks translations incommensurately to the background lattice. They are fixed in such a way that the system minimises its free energy and as we will also show, the thermodynamically preferred black holes have $k_{i}=0$. However, for our purposes it is still useful to consider the black hole backgrounds in which $k_{i}\neq 0$. Finally, we note that even though our black holes \eqref{eq:DC_ansatz} will in general break all internal $U(1)$'s and spatial translations, the combination of transformations
\begin{align}
x^{i}\to x^{i}+\xi^{i}, \quad \chi_{I}\to \chi_{I}-k_{Ii}\xi^{i},\quad \sigma_{I}\to \sigma_{I}-k_{Isi}\xi^{i}
\end{align}
is still a symmetry of our solutions.

At this point it is helpful to define the bulk field
\begin{align}\label{eq:Sdef}
S_{I}=\frac{1}{2i}\left( e^{-i(k_{Ii}x^{i}+c_{I})}Z_{I}-e^{i(k_{Ii}x^{i}+c_{I})}\bar{Z}_{I}\right)\,,
\end{align}
corresponding to the uncondensed component $\mathcal{O}_{S_{I}}$ of the boundary operator $\mathcal{O}_{Z_{I}}$ and for which it is easy to check that $\langle \mathcal{O}_{S_{I}}\rangle=0$ in the broken phase. As we will see, this operator will play an important role in our discussion of hydrodynamics in sections \ref{sec:hydro} and \ref{sec:pinning} as it will couple to the gapless mode due to the symmetry breaking in the bulk. To see this, we perform a small $U(1)$ transformation $Z_{I}\to Z_{I}(1+i\,\epsilon)$ to yield $\langle\mathcal{O}_{S_{I}}\rangle=\left| \langle \mathcal{O}_{Z_{I}}\rangle\right | \epsilon$. The technical point we would like to make at this point is that the bulk fluctuations $\delta\chi_{I}$ are intimately related to the operator $\mathcal{O}_{S_{I}}$. More specifically close to the boundary we will in general have the expansion
\begin{align}
\delta S_{I}=\phi_{I}\,\delta\chi_{I}=\zeta_{S_{I}}\,(r+R)^{\Delta_{Z_{I}}-3}+\cdots +\frac{\langle \mathcal{O}_{S_{I}}\rangle}{2\Delta_{Z_{I}}-3}\,(r+R)^{-\Delta_{Z_{I}}}+\cdots\,,
\end{align}
and $\zeta_{S_{I}}$ will be the source from the field theory point of view. This source will make its appearance again in later sections when we consider the driven hydrodynamics of our system in its broken phase.

In the IR, we demand the presence of a regular Killing horizon at $r=0$ by imposing the expansion
\begin{align}\label{nhexpbh}
U\left(r\right)&=4\pi\,T\,r+\cdots\,,\qquad V_{i}=V_{i}^{(0)}+\cdots\,,\qquad a=a^{(0)}\,r+\cdots\,,\nn
\phi_{I}&=\phi^{(0)}_{I}+\cdots\,,\qquad \psi_{I}=\psi^{(0)}_{I}+\cdots\,.
\end{align}
According to this notation, $T$ will be the Hawking temperature of the black hole horizon. In the end we will have a set of black hole backgrounds labeled by $k_{i},k_{si},\psi_{Is},\mu$, $T$ and $c_{I}$ of which all the thermodynamic charges of the system are independent.

\subsection{Thermodynamics}\label{subsec:thermodynamics}

In this section we will take the opportunity to discuss aspects of thermodynamics which will let us highlight quantities that will play a role later in our paper. In order to discuss thermodynamics we need to add appropriate counterterms $S_{bdr}$ to our bulk action \eqref{eq:bulk_action} that will make it finite and also end up with a well defined variational problem in which we will keep fixed the right sources. At leading order in their divergence in a near conformal boundary expansion, the counterterm action will include the terms\footnote{Here we are listing all the terms which are relevant for scalar operator with dimensions $\Delta \leq 9/4$. More generally there is additional terms needed to render the variational problem well posed \cite{Papadimitriou:2011qb}. Moreover, for $\Delta>5/2$ the derivative terms we have already listed in \eqref{eq:bdy_action} need to be multiplied by $U(1)^{4}$ invariant functions which depend on the complex scalars. At the order we are working in our derivative expansion the treatment of section \ref{sec:hydro} would remain valid after dropping the contribution of potential contact term contributions from one point functions.}\cite{Balasubramanian:1999re,Skenderis:2002wp}
\begin{align}\label{eq:bdy_action}
S_{bdr}=&\int_{\partial M}d^{3}x\,\sqrt{-\gamma}\,\left(-2K + 4 +R_{bdr}\right)\notag\\
&\quad-\frac{1}{2}\int_{\partial M}d^{3}x\,\sqrt{-\gamma}\,\sum_{I}\,[(3-\Delta_{Z_{I}})\bar{Z}_{I}Z_{I}+(3-\Delta_{Y_{I}})\bar{Y}_{I}Y_{I}]\notag\\
&\quad+\frac{1}{2}\int_{\partial M}d^{3}x\,\sqrt{-\gamma}\,\sum_{I}\,[\frac{1}{2\Delta_{Z_{I}}-5}\,\partial_{a}\bar{Z}_{I}\partial^{a}Z_{I}+\frac{1}{2\Delta_{Y_{I}}-5}\,\partial_{a}\bar{Y}_{I}\partial^{a}Y_{I}]+\cdots\,.
\end{align}
The counterterms are to be evaluated on a hypersurface $\partial M$ of constant holographic radius, $\gamma_{\mu\nu}$ is the induced metric on that surface and $a$ labels its coordinates. The precise form of the terms we have omitted in \eqref{eq:bdy_action}  will depend on the details of the functions that appear in our bulk action \eqref{eq:bulk_action}. However, the ingredients we will need for our analysis are not going to depend on these details.

In order to discuss thermodynamics we analytically continue to imaginary time $t=-i \tau$ and consider the renormalised Euclidean action $I_{E}=-i S_{tot}$ with $S_{tot}=S_{bulk}+S_{bdr}$. The total free energy of the system is then simply $W_{FE}=T\,I_{E}$ which is of course infinite since we are dealing with an infinite system. For our model, the free energy density $w_{FE}$ will be constant in the boundary coordinates $x^{i}$ since we are dealing with a homogeneous system and the conserved charges of the system are invariant with respect to the bulk $U(1)$ symmetries.

If $\epsilon$ is the energy density, $s$ is the entropy density, and $\rho$ is the electric charge density we have
\begin{align}
w_{FE}=\epsilon- T\,s-\mu\,\rho\,.
\end{align}
We note that our solutions are functions of $k_{i},k_{si},\psi_{Is},\mu$, $T$ and $c_I$. All the thermodynamic quantities are going to depend on all of them except for $c_{I}$. In the forthcoming derivations we will encounter the electric charge and entropy densities written in terms of the black hole horizon data as
\begin{align}\label{eq:rhos_hor}
\rho=e^{V_{1}^{(0)}+V_{2}^{(0)}}\,\tau^{(0)}\,a^{(0)},\quad s=4\pi\,e^{V_{1}^{(0)}+V_{2}^{(0)}}\,.
\end{align}
From thermodynamics we also know that $\rho=-\partial_{\mu}w_{FE}$ and $s=-\partial_{T}w_{FE}$ and therefore a variation of the free energy density with respect to the solution parameters that will matter later gives
\begin{align}
\delta w_{FE}=-\rho\,\delta\mu-s\,\delta T+ w^{i}\,\delta k_{i}\,.
\end{align}

After plugging our ansatz \eqref{eq:DC_ansatz} in our total Euclidean action $I$ and using the equations of motion we can easily show that
\begin{align}\label{eq:kvariation}
w^{i}=\partial_{k_{i}}w_{FE}=\int_{0}^{\infty}dr\,e^{V_{1}+V_{2}-2V_{i}}\,k_{i}\,\Phi_{I} \delta^{Ii}\,,
\end{align}
which is convergent as long as the scaling dimensions of our complex scalars $Z_{I}$ satisfy the unitarity bound $\Delta_{Z_{I}}>1/2$. Notice that we do not pick up any contribution from explicit variations of the counterterms with respect to $k_{i}$ in the absence of explicit sources for $\phi_{I}$ in \eqref{asymptsol}.

In the later sections the second variations of the free energy will show up in the calculation of the diffusion constants. It is useful to define the susceptibilities through
\begin{align}
\delta s&=T^{-1}c_{\mu}\,\delta T+\xi\,\delta\mu+\nu^{i}\,\delta k_{i}\,,\notag\\
\delta\rho&=\xi\,\delta T+\chi_{q}\,\delta \mu+\beta^{i}\,\delta k_{i}\,,\notag\\
\delta w^{i}&=-\nu^{i}\,\delta T-\beta^{i}\,\delta\mu+w^{ij}\,\delta k_{j}\,.
\end{align}
At this point we note that for the susceptibilities $\nu^{i}$, $\beta^{i}$ and $w^{ij}$, we would have to either take a second derivative of $w_{FE}$ or vary the bulk integral in \eqref{eq:kvariation} which is evaluated on-shell.

\section{Hydrodynamic Perturbations}\label{sec:pert}

In this section we will study perturbations of our bulk theory around the black hole backgrounds \eqref{eq:DC_ansatz} in a hydrodynamic expansion of long wavelengths. For clarity, we have split the presentation into two smaller subsections. The first one contains general statements about our perturbations which are independent of the hydrodynamic limit and which will be useful for section \ref{sec:hydro} as well. In the second one we give a description of our derivative expansion along with the final result for our diffusive modes. The interested reader can find the more technical aspects of our construction in Appendix \ref{app:hydro_expansion}.

\subsection{Perturbations}\label{sec:perturbations_setup}

To study perturbations with frequency $\omega$ and a non trivial wavenumber $q$ along the $x^{1}$ directions, we consider perturbations $\delta X$ of the background black hole solution \eqref{eq:DC_ansatz} with $\delta X\equiv\{\delta g_{tt},\delta g_{t1},\delta g_{rr},\delta g_{r1},\delta g_{ii},\delta a_{t},\delta a_{r},\delta a_{1},\delta \psi_I, \delta\sigma_{1},\delta \phi_I, \delta\chi_{1}\}$ corresponding to the longitudinal sector for perturbations with wavevectors parallel to the $x^{1}$ direction. All our functions depend on the bulk coordinates $(t,r,x^1)$. The homogeneity of the background allows us to Fourier transform along the spatial direction $x^1$ and the time $t$, leading to the separation of variables
\begin{align}\label{eq:sep_var}
\delta X(t,r,x_{1})=e^{-i\omega\, v_{EF} +i q x^{1}}\,\delta X(r)
\end{align}
where we have introduced 
\begin{align}\label{eq:vef}
v_{EF}=t+S(r)\,,
\end{align}
with $S(r)\to 0$ as $r\to\infty $ and $S(r)\to \frac{\ln r}{4\pi T}+S^{(1)}\,r+\cdots$ as we approach the horizon at $r\to0$. The advantage of introducing $v_{EF}$ comes from the fact that close to the horizon the time coordinate $t$ combines with the radial coordinate $r$ to form the ingoing Eddington-Finkelstein coordinate. This simplifies the boundary conditions we need to impose on the radial functions in order to achieve regular ingoing boundary conditions. We impose the expansions
\begin{align}\label{eq:gen_exp}
\delta g_{tt}(r)&= 4\pi T\,r\, \delta g_{tt}^{(0)}+\cdots\,,\quad
\delta g_{rr}(r)=\frac{\delta g_{rr}^{(0)}}{4\pi T\,r}+\cdots\, \,,\nn
\delta g_{t1}(r)&=\delta g_{t1}^{(0)}+r\,\delta g_{t1}^{(1)}+\cdots\,,\quad 
\delta g_{r1}(r)=\frac{\delta g_{r1}^{(0)}}{4\pi T\,r}+\delta g_{r1}^{(1)}+\cdots\,,\nn
\delta g_{ii}(r)&=\delta g_{ii}^{(0)}+\cdots\,,\quad\quad
\delta g_{tr}(r)=\delta g_{tr}^{(0)}+\cdots,\quad \delta a_{1}(r)=\delta a_{1}^{(0)}+\cdots\,,\nn
\delta a_{t}(r)&=\delta a_{t}^{(0)}+\delta a_{t}^{(1)}\,r+\cdots\,,\quad
\delta a_{r}(r)=\frac{1}{4\pi T\,r} \delta a_{r}^{(0)}+\delta a_{r}^{(1)}+\cdots\,,\nn
\delta \psi_I(r)&=\delta \psi_I^{(0)}+\cdots,\quad \delta \phi_I(r)=\delta \phi_I^{(0)}+\cdots,\quad \delta \chi_1(r)=\delta \chi_1^{(0)}+\cdots,\quad \delta \sigma_1(r)=\delta \sigma_1^{(0)}+\cdots\,,
\end{align}
which are compatible with the equations of motion. In order to achieve regularity, need to be supplemented by
\begin{align}\label{eq:nh_reg}
-2\pi T(\delta g_{tt}^{(0)}+\delta g_{rr}^{(0)})=-4\pi T\,\delta g_{rt}^{(0)}&\equiv p\,,\notag\\
\delta g_{t1}^{(0)}=\delta g_{r1}^{(0)}&\equiv-v,\notag\\
\delta a_{r}^{(0)}=\delta a_{t}^{(0)}&\equiv \varpi\,.
\end{align}
It is useful to note that at the current stage of the discussion, the fifteen constants $\delta g_{tt}^{(0)}$, $\delta g_{tt}^{(1)}$, $\delta g_{ii}^{(0)}$, $\delta a_{1}^{(0)}$, $\delta a_{t}^{(1)}$, $\delta \psi_{I}^{(0)}$, $\delta \phi_{I}^{(0)}$, $\delta \chi_{1}^{(0)}$, $\delta \sigma_{1}^{(0)}$, $\varpi$, $p$ and $v$ are constants of integration and therefore free.

Our functions $\delta X(r)$ satisfy a system of differential equations, twelve of which contain second order derivatives of our functions in the radial coordinate $r$. At the same time, we need to impose a set of four independent constraints originating from diffeomorphism and gauge invariance. In a radial foliation of spacetime by hypersurfaces orthogonal to the form $n=dr$, these constraints contain only first order derivatives in the radial coordinate $r$ and we can choose to impose them on any slice of constant $r$. The functions $\delta g_{r\mu}$ and $\delta a_{r}$ are simply Lagrange multipliers which can be chosen freely up to the boundary conditions that we gave in \eqref{eq:gen_exp} and \eqref{eq:nh_reg} and which guarantee regularity of the foliation. In more precise terms, using the notation of equation \eqref{eq:eom1} the constraints take the form $L_{\mu}=n_{\lambda}E^{\lambda}{}_{\mu}=0$ where $E_{\mu\nu}=L_{\mu\nu}-\frac{1}{2}g_{\mu\nu}L^{\rho}{}_{\rho}$ and $C=n_{\lambda}\,C^{\lambda}=0$. These can be imposed on any constant $r$ hypersurface since $\nabla_{\lambda}E^{\lambda}{}_{\mu}=0$ and $\nabla_{\lambda}C^{\lambda}=0$.

In this section we will choose the hypersurface we impose our constraints on to be infinitesimally close to the background event horizon at $r=0$. For purposes which will become more clear in section \ref{sec:hydro} we define the horizon electric and heat currents through
\begin{align}\label{eq:J_hor2}
Q_{(0)} &=4\pi T e^{V_{2}^{(0)}-V_{1}^{(0)}}v\,,\nn
J_{(0)}&=e^{V_{2}^{(0)}-V_{1}^{(0)}}\tau^{(0)}\left(iq \varpi+{a^{(0)}}v+i\omega \delta a^{(0)}_{1}\right)\,.
\end{align}
After these definitions, the aforementioned constraints can be written as \cite{Donos:2017ihe}
\begin{subequations}\label{hconstraints_all}
\begin{align}
&iq\, Q_{(0)}=i2\pi\omega Te^{V_{1}^{(0)}+V_{2}^{(0)}}\Big( e^{-2V_{1}^{(0)}}\delta g^{(0)}_{11}+e^{-2V_{2}^{(0)}}\delta g_{22}^{(0)}\Big)\,,\label{hconstrainta}\\
&iq\, J_{(0)}=i\omega e^{V_{1}^{(0)}+V_{2}^{(0)}}\Big[
\tau^{(0)}\left(a^{(0)}\,\left(\delta g^{(0)}_{tt}+\frac{p}{4\pi T}\right) +\delta a^{(1)}_{t}-\frac{i \omega}{4\pi T}\left( \delta a^{(1)}_{t}-\delta a^{(1)}_{r} \right)\right)
\nn
&\qquad+\frac{1}{2}\tau^{(0)} a^{(0)}\left(e^{-2V_{1}^{(0)}}\delta g^{(0)}_{11}+e^{-2V_{2}^{(0)}}\delta g_{22}^{(0)} \right)
+\partial_{\phi^I}\tau^{(0)}a^{(0)}\,\delta\phi^{I(0)}+\partial_{\psi^I}\tau^{(0)}a^{(0)}\,\delta\psi^{I(0)}\Big]\,,\label{hconstrainttwo}\\
&2q^{2}e^{-2V_{1}^{(0)}}\,v_{} -i\tau^{(0)}a^{(0)}\left( q\,\varpi+\omega\delta a_{1}^{(0)} \right) +iq(1+\frac{i\omega}{4\pi T})\,p \nn
&\qquad+\Psi_{1}^{(0)}k_{s1}\,\left(e^{-2V_{1}^{(0)}}k_{s1}v-i\omega \delta\sigma_{1}^{(0)} \right)+\Phi_{1}^{(0)}k_{1}\,\left(e^{-2V_{1}^{(0)}}k_{1}v-i\omega \delta\chi^{(0)}_{1} \right)\nn
&\qquad =i\omega\left(\delta g_{t1}^{(1)}
-\frac{i\,\omega}{4\pi T}(\delta g_{t1}^{(1)}-\delta g_{r1}^{(1)})
+2V_{1}^{(1)}\,v
-iq\,\delta g_{tt}^{(0)} 
-iqe^{-2V_{1}^{(0)}}\,\delta g^{(0)}_{11} \right)\,.\label{hiconstraint2b}
\end{align}
\end{subequations}
We have omitted the ``Hamiltonian'' constraint, as it is implied by those listed above.

Close to the conformal boundary, the asymptotic expansion of functions reads
\begin{align}\label{eq:gen_UVexp}
\delta g_{tt}(r)&= \mathcal{O}(r^{0})\,,\quad
\delta g_{rr}(r)=\mathcal{O}(r^{-4})\,,\quad\delta g_{t1}(r)=r^{2\,}\frac{\zeta}{i\omega}+\mathcal{O}(r^{0})\,,\quad 
\delta g_{r1}(r)=\mathcal{O}(r^{-3})\,,\nn
\delta g_{ii}(r)&=\mathcal{O}(r^{0})\,,\quad
\delta g_{tr}(r)=\mathcal{O}(r^{-2})\,,\nn
\delta a_{1}(r)&=\frac{E-\mu\,\zeta}{i\omega}\,,\quad \delta a_{t}(r)=\mathcal{O}(r^{-1})\,,\quad \delta a_{r}(r)=\mathcal{O}(r^{-2})\,,\nn
\delta \psi_I(r)&=\mathcal{O}(r^{-\Delta_{Y^{I}}})\,,\quad \delta \phi_I(r)=\mathcal{O}(r^{-\Delta_{Z^{I}}})\,,\nn
\delta \chi_1(r)&=\frac{\zeta_{S_{1}}}{\phi_{1v}}\,r^{2\Delta_{Z_{1}}-3}+\mathcal{O}(r^{2\Delta_{Z_{1}}-4})\,,\quad \delta \sigma_1(r)=\mathcal{O}(r^{-\Delta_{Y^{I}}})\,,
\end{align}
where we have included a time dependent thermal gradient source $\zeta$, external electric field $E$ and scalar source $\zeta_{S_{1}}$. Since in this section we are looking for the quasinormal modes of our black holes, we will set them equal to zero. We will switch them back on in section \ref{sec:hydro} where we will consider the driven hydrodynamics of our system. In order to complete our discussion on the systematics of our solution for the perturbation, we note that there is another twelve constants of integration that we have not listed in the expansion \ref{eq:gen_UVexp} and which are not fixed by the equations of motion. Put together with the fifteen constants of integration we have listed below the near horizon expansion \eqref{eq:gen_exp}, there is a total of twenty seven constants.

When the sources are set to zero, and nothing set sets a scale for our linear system, the equations and the boundary conditions are scale invariant and we can set any one of the constants equal to one. This suggests that we have twenty six constants to solve the twelve second order equations and the three constraints. Therefore, in the source free case we can find solutions only for discrete values of the frequency $\omega$ for a fixed wavelength $q$ which are precisely the quasinormal modes of our black hole backgrounds. In the next section we will consider the systematics of quasinormal modes which represent the hydrodynamic excitations of our system.

\subsection{Hydrodynamic modes}\label{sec:perturbations_hydro}

The hydrodynamic modes that we will consider in this section have $\omega\to 0$ as $q\to 0$ and therefore they become static, source free modes in the infinite wavelength limit. In order to understand their structure, we will construct them perturbatively by taking $q\approx \mathcal{O}(\varepsilon)$ and expanding
\begin{align}\label{eq:X_epsilon_exp}
\omega&=\varepsilon\,\omega_{[1]}+\varepsilon^{2}\,\omega_{[2]}+\cdots\notag\\
\delta X(r)&=\delta X_{[0]}(r)+\varepsilon\delta X_{[1]}(r)+\varepsilon^{2}\delta X_{[2]}(r)+\cdots\,.
\end{align}
Identifying the static source free perturbations of our black hole backgrounds \eqref{eq:DC_ansatz} is therefore a key ingredient in constructing the above hydrodynamic series. The two universal modes for a system in which translations are broken explicitly are related to thermodynamic energy and charge perturbations \cite{Donos:2017ihe}. However, for our system in which a continuous global symmetry is spontaneously broken there is an additional mode related to shifts of the constants $c_{I}$ in \eqref{eq:DC_ansatz}, the Goldstone mode.

For the thermal states with $k_{1}=0$ in \eqref{eq:DC_ansatz} the order parameter of spontaneous symmetry breaking does not break translations in the $x^{1}$ direction. In this case the perturbation $\delta\chi_{1}$ completely decouples from the rest of the system and one can sharply divide the hydrodynamic modes in the $\varepsilon\to0$ limit to the ones that have their origin in thermoelectric perturbations \cite{Donos:2017ihe} and the ones which are long wavelength excitations of the Goldstone modes\cite{Donos:2019txg}. In this paper we will consider the case with $k_{1}\neq 0$ and study in detail the mixing these two types of modes which describe different physics. In the $k_{1}=0$ case the thermoelectric fluctuations are captured by incoherent hydrodynamics. We are going to give an enlarged framework of hydrodynamics in order to capture fluctuations of the gapless mode emerging from the symmetry breaking.

The most effective way to construct the static modes associated to energy and charge fluctuations is to simply start by varying the backgrounds \eqref{eq:DC_ansatz} with respect to the temperature $T$ and the external chemical potential $\mu$. A naive perturbation variation $T\to T+ \delta T_{[0]}$ and $\mu\to \mu +\delta \mu_{[0]}$ in the functions that appear in \eqref{eq:DC_ansatz} would certainly produce solutions of the equations of motion. However, it is easy to see from the asymptotics \eqref{asymptsol} and \eqref{nhexpbh} that this would generate perturbations which are not compatible with our ingoing boundary conditions \eqref{eq:gen_exp} and \eqref{eq:nh_reg}, and moreover would introduce a boundary source for the gauge field. To remedy this, one can simply perform bulk diffeomorphisms and gauge transformations, as outlined in Appendix \ref{app:hydro_expansion}. The aim is to bring our solution \eqref{eq:DC_ansatz} in a class of coordinate systems and gauge choices such that a straightforward variation with respect to temperature and chemical potential has the desired asymptotics \eqref{asymptsol} and \eqref{nhexpbh}.

In addition to varying $T$ and $\mu$, we use the broken bulk symmetry to generate the small static shift $\delta\chi_{1}=\delta c_{g[0]}$. The resulting static solution is then
\begin{align}\label{eq:deltaX0}
\delta X_{[0]}=\frac{\partial X_{b}}{\partial T}\,\delta T_{[0]}+\frac{\partial X_{b}}{\partial \mu}\,\delta \mu_{[0]} +\frac{\partial X_{b}}{\partial c_{1}}\,\delta c_{g[0]}\,,
\end{align}
where $X_{b}$ is the transformed background according to our previous discussion. By construction, this is going to be a perturbative solution of our equations of motion at $\varepsilon=0$ and with $\delta T_{[0]}$, $\delta \mu_{[0]}$, $\delta c_{g[0]}$ independent of each other. For the case with $k_{1}=0$, we would be able to study the modes generated by the temperature $\delta T_{[0]}$ and chemical potential $\delta\mu_{[0]}$ perturbations independently from the bulk Goldstone perturbation generated by $\delta c_{g[0]}$.

When we take $\varepsilon$ to be small, the derivatives of the exponential of our total perturbation \eqref{eq:sep_var} will produce terms that are of order $\mathcal{O}(\varepsilon)$ and are specified by the functions $\delta X_{[0]}$. The resulting equations will be an inhomogeneous system of equations that $\delta X_{[1]}$ will have to satisfy. As a generalisation of \eqref{eq:deltaX0}, we can split off from $\delta X_{[n]}$ the solution $\delta \tilde{X}_{[n]}$ of the corresponding inhomogeneous system,
\begin{align}
\delta X_{[n]}=\delta \tilde{X}_{[n]}+\frac{\partial X_{b}}{\partial T}\,\delta T_{[n]}+\frac{\partial X_{b}}{\partial \mu}\,\delta \mu_{[n]}+\frac{\partial X_{b}}{\partial c_{1}}\,\delta c_{g[n]}\,.
\end{align}
Such a split is meaningful as long as we impose that the inhomogeneous piece $\delta \tilde{X}_{[n]}$ has $\tilde{p}_{[n]}=\tilde{\varpi}_{[n]}=\delta \tilde{\chi}_{1[n]}^{(0)}=0$, according to the definitions in \eqref{eq:gen_exp} and \eqref{eq:nh_reg}.

As we explain in Appendix \ref{app:hydro_expansion}, when solving the constraints \eqref{hconstraints_all} and the radial equation \eqref{eq:bulk_eom_chi} at order $\mathcal{O}(\varepsilon)$, we obtain a set of relations between $\delta T_{[0]}$, $\delta \mu_{[0]}$, $\delta c_{g[0]}$ and $\omega_{[1]}$. That system of equations gives that $\delta T_{[0]}=\delta \mu_{[0]}=\omega_{[1]}=0$ as long as $k_{1}\neq 0$ such that the order parameter breaks translations. We therefore see that temperature and chemical potential perturbations will mix at higher order in $\varepsilon$ with the spatially dependent Goldstone mode. This is intuitively expected, since the constant $\delta c_{g[0]}$ only shifts the Goldstone mode, the system is going to be energetically affected only through its gradient which is of order $\mathcal{O}(\varepsilon)$. We therefore expect that the variations with respect to the temperature and the chemical potential will start mixing at order $\mathcal{O}(\varepsilon)$ and the first non-zero contributions will be $\delta T_{[1]}$ and $\delta\mu_{[1]}$.

A further point which lets us make progress in Appendix \ref{app:hydro_expansion} is the observation that after setting $\delta T_{[0]}=\delta \mu_{[0]}=\omega_{[1]}=0$ in \eqref{eq:deltaX0}, we can think of the approximation
\begin{align}
\delta \chi_{1}(t,x^{1})\approx \delta c_{g[0]}+i q\, x^{1}\, \delta c_{g[0]}+\mathcal{O}(\varepsilon^{2})\,,
\end{align}
for any finite value of $x^{1}$. This is telling us that at order $\mathcal{O}(\varepsilon)$, all that $\delta c_{g[0]}$ does is the shifts $c_{I}\to c_{I}+ \delta c_{g[0]}$ and $k_{1}\to k_{1}+i\,q\,\delta c_{g[0]}$ in \eqref{eq:DC_ansatz}. We therefore conclude that $\delta \tilde{X}_{[1]}$ has to be such that when we expand the full pertubation \eqref{eq:sep_var} at $\mathcal{O}(\varepsilon)$, we will obtain a perturbation of the background $X_{b}$ respect to $k_{1}$. As we just saw, the part of the perturbation containing the charged fields under the bulk $U(1)$'s is already contained in $\delta X_{[0]}$. This suggests that $\delta \tilde{X}_{[1]}$ can only contain the variation of the background fields $X_{b}^{N}$ which are neutral under the $U(1)$'s. More generally we found it useful to further split the $n$-th solution $\delta \tilde{X}_{[n]}$ of the inhomogeneous systems according to
\begin{align}
\delta \tilde{X}_{[n]}=\delta \mathbb{X}_{[n]}+iq\,\frac{\partial X^{N}}{\partial k_{1}}\,\delta c_{g[n-1]}\,,
\end{align}
and according to our discussion we have $\delta \mathbb{X}_{[1]}=0$.

Finally, in Appendix \ref{app:hydro_expansion} we examine the radial equation \eqref{eq:bulk_eom_chi} at order $\mathcal{O}(\varepsilon^{2})$ and the constraints \eqref{hconstraints_all} at order $\mathcal{O}(\varepsilon^{3})$. This gives us a homogeneous system of linear equations that the constants $\delta T_{[1]}$, $\delta\mu_{[1]}$ and $\delta c_{g[0]}$ have to satisfy. Written in a matrix form, the system reads
\begin{align}\label{3by3}
\left(\mathbf{X}_H -\mathbf{\Sigma}_H\right) \begin{pmatrix}
q\,\delta c_{g[0]} \\
\delta T_{[1]} \\
\delta \mu_{[1]}
\end{pmatrix}=0
\end{align}
where
\begin{align}\label{xmatrix}
\mathbf{X}_H\equiv 
\left(\begin{array}{ccc} 
- iq^2\,w^{11} & q^{2}\,\nu^1 & q^{2}\,\beta^1 \\ 
-\omega_{[2]}\, \nu^1 & i\omega_{[2]}T^{-1}c_\mu & i\omega_{[2]}\xi \\
-\omega_{[2]}\, \beta^1 & i\omega_{[2]}\xi & i\omega_{[2]}\chi_q \\
\end{array}\right)\,,
\end{align}
and
\begin{align}\label{sigmamatrix}
\mathbf{\Sigma}_H\equiv
\left(\begin{array}{ccc}
\omega_{[2]}\,\left(\bar{\vartheta}-\varrho_H \bar{\lambda}\right) & q^{2}\,T^{-1}\lambda_H & q^{2}\,\gamma_H \\
-\omega_{[2]}\,T^{-1}\lambda_H & q^2 T^{-1}\bar{\kappa}_H & q^2 \bar{\alpha}_H \\
-\omega_{[2]}\,\gamma_H & q^2 \alpha_H & q^2 \sigma_H \\
\end{array}\right)\,.
\end{align}
In the above expressions we are using the notation of subsection \ref{subsec:thermodynamics} along with the definitions
\begin{align}\label{eq:horizon_coeffs}
\alpha_{H}&=\bar\alpha_{H}= \frac{4 \pi \rho}{\mathcal{B}}\,,\quad
\sigma_{H}=\frac{s \, e^{-2V_{1}^{(0)}} \, \tau^{(0)}}{4 \pi} + \frac{4 \pi \rho^2}{s\, \mathcal{B}}\,,\nn
\bar\kappa_{H}&=\frac{4 \pi T s}{\mathcal{B}}\,,\quad
\lambda_{H}= \frac{k_1 \Phi_1^{(0)} T s}{\mathcal{B}}\,,\quad
\gamma_{H}=\frac{k_1 \Phi_1^{(0)} \rho}{\mathcal{B}}\,,\notag\\
\bar{\vartheta} &= \frac{\Phi_1^{(0)}}{4\pi T}\left(T s +k_{1}w^1 \right)\,,\quad \varrho_H = \frac{k_{1}\Phi_1^{(0)}}{4\pi T}\,,\quad
\bar\lambda = \lambda_H + w^1\,,\nn
\mathcal{B}&=k_{1}^{2}\,\Phi^{(0)}_{1}+k_{s1}^{2}\,\Psi^{(0)}_{1}\,.
\end{align}
In order for the linear system \eqref{3by3} to have non-trivial solutions, we must demand that the matrix of coefficients is non-invertible. The vanishing of the determinant of $\mathbf{X}_H -\mathbf{\Sigma}_H$ then fixes the dispersion relations of the three modes we are after. It is clear from the form of the matrices $\mathbf{X}_{H}$ and $\mathbf{\Sigma}_{H}$ that we obtain three diffusive modes of the form
\begin{align}\label{eq:diff_dispersion}
\omega_{i}=-i\,D_{i}\,q^{2},\quad i=1,2,3\,,
\end{align}
with diffusion constants $D_{i}$ expressed in terms of thermodynamic susceptibilities and the coefficients in \eqref{eq:horizon_coeffs}. In section \ref{sec:hydro}, we will derive a hydrodynamic theory which precisely reproduces these modes. There, we will find the quantities \eqref{eq:horizon_coeffs} appearing as transport coefficients in the constitutive relations for the currents along with a Josephson-type relation for the gapless mode of the spontaneous breaking.

Here we note that setting $k_{1}=0$ gives $\lambda_{H}=\gamma_{H}=\beta^{1}=\nu^{1}=0$, bringing the matrix of coefficients in \eqref{3by3} in a block diagonal form. This demonstrates the decoupling between the thermoelectric and the bulk Goldstone modes. In this limit, the coefficients $\sigma_{H}$, $\alpha_{H}$, $\bar{\alpha}_{H}$ and $\bar{\kappa}_{H}$ coincide with the DC thermoelectric transport coefficients of the boundary theory. The relevant diffusion constants then satisfy a generalised version of Einstein's relations \cite{Hartnoll:2014lpa,Donos:2017gej,Donos:2017ihe}. This makes clear that the extra diffusive mode that appears in our theory has nothing to do with the spontaneous breaking of translations, it describes the same physics with the setup of \cite{Donos:2019txg}. At finite $k_{1}$ though, we see that the two different types of modes mix with each other. This will become much clearer in the next sections where we give a hydrodynamics description and we include external sources and a gap. In this framework, one can also use the standard formalism of hydrodynamics in order to derive the linear system of equations \eqref{3by3} which fix the dispersion relations of the diffusive modes.

\section{Incoherent hydrodynamics and density waves}\label{sec:hydro}
In this section we wish to derive a theory of hydrodynamics which captures the physics of long wavelength excitations in our system. In the infinite wavelength limit, we have seen that our gapless modes describe fluctuations in temperature, the chemical potential and phase shifts for the dual operators of the bulk fields $Z_{I}$. In order to give a complete description of the system, we need to identify the correct conservation laws and effective description of the bulk Goldstone mode.

The conserved currents we will be interested in are the electric current $J^{\mu}$ associated to charge conservation and the heat current $Q^{\mu}$ we can construct in perturbation theory associated to time translations of the background \eqref{eq:DC_ansatz}. To see how this works we note that the global $U(1)$ and diffeomorphism symmetries of the boundary theory imply the current and stress tensor $T_{\mu\nu}$ Ward identities
\begin{align}
\nabla_{a}J^{a}&=0\notag\\
\nabla_{a}T^{a}{}_{b}&=F_{b a}J^{a}+\frac{1}{2}\left(\nabla_{b}\bar Y_{Is}\mathcal{O}_{Y_{I}}+\nabla_{b}\bar Z_{Is}\mathcal{O}_{Z_{I}}+\mathrm{c.c.}\right)\,,
\end{align}
with $F=dA$ the field strength of the external source one-form $A_{a}$ and $\bar Y_{Is}$, $\bar Z_{Is}$ are the sources for the complex scalar operators. Contracting the stress tensor Ward identity with a vector $k^{\mu}$ gives
\begin{align}
\nabla_{a} \left[\left(T^a{}_b + A_b J^a \right)k^b\right] =\frac{1}{2}T^{ab}\,\mathcal{L}_k g_{ab} +J^{a}\mathcal{L}_k A_{a} +\frac{1}{2}\left(\mathcal{L}_k \bar Y_{Is}\mathcal{O}_{Y_{I}}+\mathcal{L}_k\bar Z_{Is}\mathcal{O}_{Z_{I}}+\mathrm{c.c.}\right)\,.\notag
\end{align}
In contrast to section \ref{sec:pert}, we will add the thermal gradient $\zeta$ and electric field $E$ perturbations which will enter the boundary metric $g_{ab}$ and external field $A_{a}$ according to
\begin{align}\label{eq:sources}
\delta ds^{2}&=2\,(i\omega)^{-1}\,\zeta\,e^{-i\omega\,t+i q x^{1}}\,dt\,dx^{1},\quad \delta A=(i\omega)^{-1}\,(E-\mu\,\zeta)\,e^{-i\omega\,t+i q x^{1}}\,dx^{1}\,,
\end{align}
along with the source $\delta Z_{1s}$ for the scalar field
\begin{align}\label{eq:sources2}
\delta Z_{1s}&=\frac{i}{2}\,e^{i(k_{1}x^{1}+c_{1})}\,\zeta_{S_{1}}\,e^{-i\omega\,t+i q x^{1}}\,.
\end{align}
We are now going to make the choice $k=\partial_{t}$ and perturbatively expand the contracted Ward identity to give the electric current and heat conservation
\begin{align}\label{eq:conservation_laws}
\partial_{a} \delta J^{a}=&0\notag\\
\partial_{a}\delta Q^{a}=&0
\end{align}
with $\delta Q^{a}=-\delta T^{a}{}_{t}-\mu\,\delta J^{a}$.

In order to obtain a closed system of equations, apart from the conservations laws \eqref{eq:conservation_laws} we need two additional ingredients. The first is to express the boundary theory currents $\delta J^{a}$ and $\delta Q^{a}$ in a derivative expansion of the local variations $\delta \hat{\mu}(t, x^{1})$, $\delta \hat{T}(t, x^{1})$ and $\delta \hat{c}_{g}(t, x^{1})$. At leading order in our derivative expansion we identify them as the Fourier modes
\begin{align}\label{eq:hydro_var}
\delta\hat{\mu}=e^{-i\omega t + iq x^1}\delta\mu_{[1]}\,,\qquad \delta\hat{T}=e^{-i\omega t + iq x^1}\delta T_{[1]}\,,\qquad \delta\hat{c}_g=e^{-i\omega t + iq x^1}\delta c_{g[0]}\,.
\end{align}
The second ingredient is to find an effective description for the dynamics of the phase $ \delta\hat{c}_g$. Following closely the techniques of \cite{Donos:2019txg}, in our holographic model this is going to come from correctly identifying the sources for the field theory dual of $S_{I}$ as defined in \eqref{eq:Sdef}. The physical interpretation of $\delta \hat{c}_{g}(t, x^{1})$ comes after reminding the reader that at leading order in epsilon we have
\begin{align}\label{eq:SVEV_hydro}
\langle \mathcal{O}_{S_{1}}\rangle = \left| \langle \mathcal{O}_{Z_{1}}\rangle\right|\,\delta\hat{c}_{g}\,.
\end{align}

Since we are going to study holographic models, it is useful to note that the continuity equations \eqref{eq:conservation_laws} are equivalent to the constraints $L_{\mu}k^{\mu}=0$ and $C=0$ when evaluated at infinity with $k=\partial_{t}$. At this point we see that the philosophy of this section is going to be slightly different from that of section \ref{sec:pert} and Appendix \ref{app:hydro_expansion}. As we explained there, the system of the final equations \eqref{3by3} that fixed the dispersion relations, is the constraints which we chose to impose on a hypersurface close to the black hole horizon \eqref{hconstraints_all} along with the equation of motion \eqref{eq:bulk_eom_chi}. Of course we had to make sure that all our other radial equations admitted a solution and this was guaranteed by the way we constructed our $\varepsilon$-expansion.

Here we will choose to impose the Gauss and time component $L_{t}$ of the momentum constraints on a constant $r$ surface at infinity. The other components $L_{\mu}$ of the momentum constraints with $\mu\neq t$ will still be imposed close to the horizon, just as we did in section \ref{sec:pert}. This is in general not possible since all the momentum constraints need to be imposed at the same hypersurface. We therefore need to show that $\partial_{r}(\delta L_{t})=0$  independently of the other constraints. For any vector in the bulk $\xi^{\mu}$ we have that
\begin{align}
\nabla_{\mu}(E^{\mu}{}_{\nu}\xi^{\nu})=\nabla_{(\mu}\xi_{\nu)}E^{\mu\nu}\,.
\end{align}
We therefore see that if $\xi^{\mu}$ is a Killing vector for the background and the background satisfies Einstein's equations, we must have
\begin{align}
\nabla_{\mu}(\delta E^{\mu}{}_{\nu}\,\xi^{\nu})=0\,.
\end{align}
at leading order in perturbation theory from where we see that $\partial_{r}(\delta L_{t})=0$ as long as $\delta E^{a}{}_{\nu}\xi^{\nu}=0$ with $a\neq r$ are satisfied. Moreover, if $n_{\mu}\xi^{\mu}=0$, then we have that all of $\delta E^{a}{}_{\nu}\xi^{\nu}=0$ that we need to impose on the hypersurface are just a linear combination of the second order in $r$ equations of motion which should be imposed everywhere in the bulk. Therefore, for such a $\xi=\partial_{t}$ we have that $\partial_{r}(\delta E^{r}{}_{t})=0$ independently of the other constraints being satisfied on the hypersurface.

The above argument shows that in our situation we are allowed to independently impose the momentum constraints \eqref{hiconstraint2b} on the horizon which is a very efficient way to integrate out the horizon fluid velocity $v$ at the energy scales we are interested in.

Before making this step, it is now a good point to describe how we are going to turn on the external sources \eqref{eq:sources} in the bulk. For the electric field and temperature gradient, the most efficient to do this is to add the zeroth order terms
\begin{align}\label{eq:ezeta_sources}
\delta g_{t1}&=\cdots+e^{-i\omega v_{EF}+i q x^{1}}\left(\frac{\zeta\, U}{i \omega}\right)\,,\notag\\
\delta a_{1}&=\cdots+e^{-i\omega v_{EF}+i q x^{1}}\left(\frac{ E}{i\omega} -\frac{a\,\zeta}{i \omega}\right)\,,
\end{align}
to the discussion of section \ref{sec:pert}. The source for the complex scalar will appear later in our analysis when implementing the boundary conditions \eqref{eq:gen_UVexp}. This will happen at second order in the $\varepsilon$-expansion as $\zeta_{S_{1}}\sim \mathcal{O}(\varepsilon^{2})$. One can see that these extra terms are regular on the black hole horizon and that they correctly introduce the sources according to \eqref{eq:gen_UVexp}. Moreover, they automatically satisfy all the equations of motion up to second order in $\varepsilon$ if we take $\zeta$ and $E$ to scale like $\mathcal{O}(\varepsilon^{2})$. The easiest way to see this is to also perform the regular coordinate and gauge transformations given by
\begin{align}
&t\to t-\frac{\zeta}{(i\omega)(iq)}e^{-i\omega v_{EF}+i q x^{1}}\,,\\
& A\to A+d\delta \Lambda, \qquad \delta\Lambda=-\frac{\zeta}{(i\omega)(iq)}e^{-i\omega v_{EF}+i q x^{1}}\,.
\end{align}
The resulting perturbation is then of order $\varepsilon$ and it trivially satisfies the equations of motion up to order $\mathcal{O}(\varepsilon)$. To see this one needs to just strip off the oscillating exponential and notice that after these transformations, the new perturbative terms are just a rescaling of the time coordinate in \eqref{eq:DC_ansatz} and the addition of a regular exact form to the background gauge field. This shows that it is only terms coming from derivatives of the exponentials that will violate the equations of motion.

The interested reader can see how the vector constraint is modified by the sources in Appendix \ref{app:hydro}. An important ingredient we import from section \ref{sec:pert} and which enters our analysis, is that we should take the scaling $\delta\hat{c}_{g}\sim\mathcal{O}(1)$, $\delta\hat{\mu},\,\delta \hat{T},\, q\sim\mathcal{O}(\varepsilon)$ and $\omega \sim \mathcal{O}(\varepsilon^{2})$. It is worth mentioning that the necessity for these scalings can be derived using the formalism of this section. The other important step we chose to focus there is to show that the boundary electric and heat currents can be expressed as
\begin{align}
\delta J^1 &=\sigma_H \left(\hat{E} -\partial_1 \delta\hat{\mu}\right) + \alpha_H \left(T\hat{\zeta} -\partial_1 \delta\hat{T}\right) - \gamma_H\, \partial_t \delta\hat{c}_g\,,\label{eq:J_constitutive}\\
\delta Q^1& =T\,\bar{\alpha}_H \left(\hat{E} -\partial_1 \delta\hat{\mu}\right) +\bar{\kappa}_H\left(T\hat{\zeta} -\partial_1 \delta\hat{T}\right) -\bar{\lambda}\, \partial_t \delta\hat{c}_g\,,\label{eq:Q_constitutive}
\end{align}
with the transport coefficients exactly as defined in \eqref{eq:horizon_coeffs}. From the above expressions we see that the currents themselves are of order $\mathcal{O}(\varepsilon^{2})$ implying that we should take the charge densities $\delta J^{t}$ and $\delta Q^{t}$ up to order $\mathcal{O}(\varepsilon)$ in order to solve the constraints \eqref{eq:conservation_laws} up to order $\mathcal{O}(\varepsilon^{3})$. Remembering the structure of our derivative expansion \eqref{eq:X_epsilon_exp}, we see that the zeroth order perturbation $\delta X_{[0]}$ does not have an effect on the thermodynamic quantities of our system. The first non-trivial corrections come from the first correction $\delta X_{[1]}$ which simply gives
\begin{align}\label{eq:charge_variation}
\delta J^{t}&=\delta \rho=\xi \delta\hat{T} +\chi_q \delta\hat{\mu} +\beta^1\partial_{1}\delta\hat{c}_g\,,\notag\\
\delta Q^{t}&=-\delta T^{t}{}_{t}-\mu\,\delta J^{t}=\delta\epsilon-\mu\,\delta\rho= c_\mu \delta\hat{T} +T\xi\, \delta\hat{\mu} +T\nu^1\partial_{1}\delta\hat{c}_g+w^{1}\,\partial_{1}\delta\hat{c}_g\,.
\end{align}
The equations we would then get from \eqref{eq:conservation_laws} are equivalent to the ones we would get from
\begin{align}\label{eq:conservation_laws_v2}
\partial_{t}\hat{\rho}+\partial_{1}\delta J^{1}=&0\,,\\
T\,\partial_{t}\hat{s}+\partial_{1}\left(\delta Q^{1}+w^{1}\partial_{t}\delta\hat{c}_{g} \right)=&0\,,
\end{align}
where we have defined the hatted thermodynamic quantities as e.g. $\hat{\rho}\equiv\rho(\mu+\delta\hat{\mu},T+\delta\hat{T},k_{1}+\partial_{1}\delta\hat{c}_{g})$.

Finally, we need to state the Josephson-type equation which fixes the time derivative of $\delta\hat{c}_{g}$. This can be simply obtained by following the treatment of Appendix \ref{app:hydro}  and in particular from the asymptotics of the solution of $\delta\chi_{1[2]}$ in equation \eqref{eq:chi12_as_exp}. In combination with the asymptotic expansion for the background field $\phi_{1}$ we can identify the source $\zeta_{S}$
\begin{align}\label{eq:dcg_eom_q}
&\bar{\vartheta}\, \partial_t\delta\hat{c}_{g} +\varrho_H\, \delta Q^1 -\,\partial_1 \hat{w}^1 +w^1\,\hat{\zeta}= \left | \langle \mathcal{O}_{Z_{1}}\rangle \right |\,\hat{\zeta}_{S_{1}}\,,
\end{align}
where we have used the boundary expression for the heat current \eqref{eq:bdy_hor_heat} to eliminate $v_{[2]}$ and with the relevant transport coefficients as defined in \eqref{eq:horizon_coeffs}. 

In order to verify that we are reproducing the same diffusive modes with section \ref{sec:pert}, we now set the sources $\hat{E}$, $\hat{\zeta}$ and $\hat{\zeta}_{S_{1}}$ to zero. It is a simple matter to check that the conservation laws \eqref{eq:conservation_laws_v2} along with the constitutive relations \eqref{eq:J_constitutive} and \eqref{eq:Q_constitutive} and the Josephson relation \eqref{eq:dcg_eom_q} reproduce the linear system of equations \eqref{3by3}. Since we have kept the sources in our description, we could also compute the AC thermoelectric conductivities of our system. We will postpone this until the end of the next section where we will also introduce a pinning parameter which relaxes the phase $\delta \hat{c}_{g}$. The aim will be to give a quantitative explanation of the AC conductivities of the setup of \cite{Donos:2019tmo} up to frequencies set by the scale of the gap.

\section{Pinning and AC transport}\label{sec:pinning}

In this section we will introduce a pinning parameter $\delta \phi_{1s}$ which adds a small explicit breaking to the global $U(1)$ associated to $Z_{1}$ in the case where its dual $\mathcal{O}_{Z_{1}}$ is not irrelevant with $\Delta_{Z_{1}}\leq 3$. This will modify the expansion of the bulk field $\phi_{1}$ close to the conformal boundary according to
\begin{align}\label{eq:phias_pinning}
\phi_{1}=  \delta\phi_{1s}\,r^{\Delta_{Z_{1}}-3}+\cdots+ \phi_{1v}\,r^{-\Delta_{Z_{1}}}+\cdots\,.
\end{align}

This small pinning parameter introduces a small gap to one of the diffusive modes we studied in sections \ref{sec:pert} and \ref{sec:hydro}. In order to quantitatively extract its effects on the physics at long wavelengths, we will incorporate it in the hydrodynamic description we discussed in section \ref{sec:hydro}. We will follow closely the discussion of \cite{Donos:2019txg} in order to do this and we will take $\delta\phi_{1s}$ to be of order $\mathcal{O}(\varepsilon^{2})$. At the order we are working, the only effect of the pinning parameter will be to modify \eqref{eq:dcg_eom_q} which is essentially the identification of the sources for $\mathcal{O}_{S_{1}}$.

At second order in $\varepsilon$, the solution of the bulk equations of motion for our fields and in particular of \eqref{eq:bulk_eom_chi} remains the same with what we had in sections \ref{sec:pert} and \ref{sec:hydro}. However, as in the previous section, the correct interpretation of the sources comes from examining the asymptotics of $\phi_{1}\,\delta\chi_{1}$ after having introduced the perturbative background source $\delta\phi_{1s}$. At the order we are working in $\varepsilon$, the asymptotics of $\delta\chi_{1}$ is still given by \eqref{eq:chi12_as_exp} where we once again substitute $v_{[2]}$ from \eqref{eq:bdy_hor_heat}. We also note that equation \eqref{eq:SVEV_hydro} still holds for the VEV of $\mathcal{O}_{S_{1}}$ at the order we are working in $\varepsilon$. Identifying the source for $S_{1}$ we find
\begin{align}\label{eq:dy_source}
&\bar{\vartheta}\, (\Omega\, \delta \hat{c}_{g} + \partial_t\delta\hat{c}_{g}) +\varrho_H\, \delta Q^1 -\,\partial_1 \hat{w}^1 +w^1\,\hat{\zeta}=\left | \langle \mathcal{O}_{Z_{1}}\rangle \right |\,\hat{\zeta}_{S_{1}}\,,
\end{align}
where we have defined
\begin{equation}\label{eq:Omega_def}
\Omega=\frac{ \left(2\Delta_{Y_1}-3\right)\phi_v }{\bar{\vartheta} }\delta \phi_{1s}=\frac{\left | \langle \mathcal{O}_{Z_{1}}\rangle \right |}{\bar{\vartheta} }\delta \phi_{1s}\,.
\end{equation}
As one might had expected, after introducing the pinning parameter $\delta\phi_{1s}$, there is a restoring force for the the phase of the complex scalar VEV $c_{1}$ which wants to bring it back to its thermal phase value. We see that $\Omega$ plays the role of a phase relaxation time but it is not quite equal to the gap of the would be diffusive mode. In order to find the gap $\omega_{g}$ we look for an exponentially decaying mode of our hydrodynamics by writing
\begin{align}
\delta\hat{T}=z_{T}\,e^{-\omega_{g}\,t}\,,\quad \delta\hat{\mu}=z_{\mu}\,e^{-\omega_{g}\,t}\,,\quad \delta\hat{c}_{g}=z_{g}\,e^{-\omega_{g}\,t}\,,
\end{align}
and setting all the sources to zero. We find that we can have a non-trivial solution for \eqref{eq:dy_source} with
\begin{equation}\label{eq:gap}
\omega_g =\frac{\bar{\vartheta}\, \Omega}{\bar{\vartheta}-\varrho_H \bar{\lambda}} =\frac{\mathcal{B} \left | \langle \mathcal{O}_{Z_{1}}\rangle \right |}{\sqrt{g_{(0)}}\,k_{s1}^2\Psi_1^{(0)}\Phi_1^{(0)}}\, \delta \phi_{1s}\,.
\end{equation}

If we were looking for a spatially dependent mode, we would find the spectrum \eqref{eq:diff_dispersion} with one of the modes acquiring a gap e.g.
\begin{align}
\omega_{1}=-i\,\omega_{g}-i\,D_{1}\,q^{2}\,.
\end{align}
Note that there exists another gapped mode in the system corresponding to the momentum relaxation pole. This mode has a gap which is much larger than the characteristic scales of the fluctuations which are captured by our hydrodynamics and  therefore does not show up in our system.

Apart from the interesting dynamics, this energy scale would show up in finite frequency transport experiments. One can think of it as the energy scale at which the density wave will be activated and contribute to transport. In order to demonstrate this we will now compute the AC transport coefficients by turning on the sources for the temperature gradient $\hat{\zeta}=e^{-i\omega t}\zeta$, electric field $\hat{E}=e^{-i\omega t}E$ and scalar source $\hat{\zeta}_{S_{1}}=e^{-i\omega t}\zeta_{S_{1}}$. It is easy to see that in this situation our hydrodynamics can be solved by simply setting $\delta\hat{T}=\delta\hat{\mu}=0$ and $\delta\hat{c}_{g}=z_{g}\,e^{-i\omega t}$. After eliminating $z_{g}$ from the currents \eqref{eq:J_constitutive} and \eqref{eq:Q_constitutive} as well as from the VEV \eqref{eq:SVEV_hydro} by using \eqref{eq:dy_source} we obtain
\begin{align}\label{eq:AC_cond}
\sigma\left(\omega\right) &=(i\omega)^{-1}\,G_{JJ}(\omega,0)= \sigma_H +\frac{T \bar{\alpha}_H \gamma_H \varrho_{H}}{\bar{\vartheta}-\varrho_H \bar{\lambda}} \frac{\omega}{\omega+i\omega_g} \,,\nn
T\alpha\left(\omega\right) &=(i\omega)^{-1}\,T\,G_{JQ}(\omega,0)= T\alpha_H +\frac{\gamma_H\left(T \bar{\kappa}_H\varrho_{H} + w^1\right)}{\bar{\vartheta}-\varrho_H \bar{\lambda}} \frac{\omega}{\omega+i\omega_g}\,,\nn
T\bar\alpha\left(\omega\right) &=(i\omega)^{-1}\,T\,G_{QJ}(\omega,0)= T\bar{\alpha}_H +\frac{T\bar{\alpha}_H \bar{\lambda}\,\varrho_{H}}{\bar{\vartheta}-\varrho_H \bar{\lambda}} \frac{\omega}{\omega+i\omega_g}\,,\nn
T\bar{\kappa}\left(\omega\right) &=(i\omega)^{-1}\,T\,G_{QQ}(\omega,0)= T\bar{\kappa}_H +\frac{\bar{\lambda}\left(T \bar{\kappa}_H\varrho_{H} + w^1\right)}{\bar{\vartheta}-\varrho_H \bar{\lambda}} \frac{\omega}{\omega+i\omega_g}\,,\nn
G_{JS}(\omega,0)&=-\frac{\left| \langle \mathcal{O}_{Z_{1}}\rangle\right|\,\gamma_{H}}{\bar{\vartheta}-\varrho_H \bar{\lambda}} \frac{\omega}{\omega+i\omega_g},\quad
G_{QS}(\omega,0)=-\frac{\left| \langle \mathcal{O}_{Z_{1}}\rangle\right|\,\bar{\lambda}}{\bar{\vartheta}-\varrho_H \bar{\lambda}} \frac{\omega}{\omega+i\omega_g}\,,\nn
G_{SJ}(\omega,0)&=\frac{\left| \langle \mathcal{O}_{Z_{1}}\rangle\right|\,T\,\bar{\alpha}_{H}\varrho_{H}}{\bar{\vartheta}-\varrho_H \bar{\lambda}} \frac{\omega}{\omega+i\omega_g},\quad
G_{SQ}(\omega,0)=\frac{\left| \langle \mathcal{O}_{Z_{1}}\rangle\right|\,(T\bar{\kappa}_{H}\varrho_{H}+w^{1})}{\bar{\vartheta}-\varrho_H \bar{\lambda}} \frac{\omega}{\omega+i\omega_g}\,,\nn
G_{SS}(\omega,0)&=\frac{\left| \langle \mathcal{O}_{Z_{1}}\rangle\right|^{2}}{\bar{\vartheta}-\varrho_H \bar{\lambda}} \frac{i}{\omega+i\omega_g} \,.
\end{align}
Here we note that since $J$ and $Q$ are odd and $S$ is even under time reversal, our retarded Green's functions have to satisfy the Onsager relations $G_{JQ}(\omega,0)=G_{QJ}(\omega,0)$, $G_{SJ}(\omega,0)=-G_{JS}(\omega,0)$ and $G_{SQ}(\omega,0)=-G_{QS}(\omega,0)$. In general this would put constraints on the transport coefficients in our theory of hydrodynamics. However, since our theory is coming from a consistent framework these are guaranteed by the specific form of our transport coefficients \eqref{eq:horizon_coeffs}. Note also that, for $k_1=0$,  $G_{SS}$ matches exactly the result of \cite{Donos:2019txg} for $q=0$. In that case we also have $G_{SJ}=G_{SQ}=0$ demonstrating the decoupling of the diffusive phase mode from the transport currents of the system.

As we might had expected, since the sliding mode couples to the heat and electric currents, the gap appears as a pole in the Green's functions relevant to transport properties. At low frequencies $\omega\ll\omega_{g}$ the sliding mode is fully pinned and all transport happens through incoherent processes and momentum relaxation in the system. In other words, by keeping $\omega_g\neq0$ and taking $\omega\to0$ we reduce to the case studied in \cite{Banks:2015wha}, with the DC conductivities given by the horizon ``conductivities'', because we have gapped the bulk Goldstone mode that couples to the heat current. An equivalent way to think about this by observing that when $\Omega\neq 0$ and the sliding mode is gapped, for frequencies $\omega\ll\omega_{g}$ one can integrate out $\delta \hat{c}_{g}$ by using \eqref{eq:dy_source}. At such frequencies, the sliding mode is a higher derivative effect in the constitutive relations \eqref{eq:J_constitutive}, \eqref{eq:Q_constitutive} and \eqref{eq:charge_variation}. On the other hand, by taking $\omega \gg \omega_{g}$ we are fully exciting the sliding mode and we see its effects of the transport properties of our thermal state; one can think of this corresponds as a frequency dependent depinning of the density wave. This is equivalent to first taking $\omega_g\to0$ in the above formulas reducing to the results of \cite{Donos:2019tmo}, which included the effects of the sliding mode already at zero frequency.

Another point that comes out of the form of the Green's functions \eqref{eq:AC_cond} is that even though there is a pole which is parametrically close to the origin at $\omega=-i\omega_{g}$, the thermoelectric transport coefficients can be arbitrarily small. The reason for this is that the effective light degree of freedom responsible for the pole $\delta \hat{c}_{g}$ couples only through its time derivative to the transport currents in \eqref{eq:J_constitutive} and \eqref{eq:Q_constitutive}. This coupling gives a residue which is parametrically small, putting the overall contribution of the degree of freedom at the same level with the diffusive terms of the local temperature and chemical potential.

\section{Numerical Checks}\label{sec:numerics}
The aim of this section to is perform numerical checks on the results of section \ref{sec:pert} and \ref{sec:hydro}. To achieve this, we need to specify the precise action we will be working with and construct the thermal states of interest. Following \cite{Donos:2019tmo}, we consider a four-dimensional Einstein-Maxwell theory coupled to six real scalars, $\phi$, $\psi$, $\chi_i$ and $\sigma_i$ with $i=1,2$,
\begin{align}\label{eq:bulk_action3}
S&=\int d^4 x \sqrt{-g}\,\Bigl(R+V(\phi)-\frac{3}{2}\left(\partial\phi \right)^2 -\frac{1}{2}\left(\partial\psi \right)^2 -\frac{1}{2}\theta(\phi)\left[\left(\partial\chi_{1} \right)^2 +\left(\partial\chi_{2} \right)^2 \right]\nn
&\qquad\qquad\qquad  -\frac{1}{2}\theta_{1}(\psi)\left[\left(\partial\sigma_{1} \right)^2 +\left(\partial\sigma_{2} \right)^2 \right] -\frac{\tau(\phi, \psi)}{4}\,F^{2} \Bigr) \,,
\end{align}
where 
\begin{align}\label{model}
&V(\phi, \psi)=-6 \cosh \phi\,,\nonumber\\
&\theta(\phi)=12 \sinh^2(\delta\, \phi)\,,\nonumber\\
&\theta_1(\psi)=\psi^2\,,\nonumber\\
&\tau(\phi, \psi)=\cosh(\gamma\, \phi)\,.
\end{align}
Although this model would appear to be outside the class \eqref{eq:bulk_action3} in the main text, it is related to them by a field redefinition of $\phi$. The variation of the above action gives rise to the following field equations of motion
\begin{align}\label{eq:eom}
&R_{\mu\nu}+\frac{\tau}{2} (F_{\mu\rho}F_{\nu}{}^{\rho}-\frac{1}{4}g_{\mu\nu}F^2)-\frac{1}{2}g_{\mu\nu} V-\frac{3}{2}\partial_\mu\phi\partial_\nu\phi-\frac{1}{2}\partial_\mu\psi\partial_\nu\psi\notag\\
&\quad-\sum_{i}(\frac{\theta}{2}\partial_\mu\chi_i\partial_\nu\chi_i+\frac{\theta_1}{2}\partial_\mu\sigma_i\partial_\nu\sigma_i)=0\,,\nonumber\\
& \frac{3}{\sqrt{-g}}\partial_\mu\left(\sqrt{-g}\,\partial^\mu\phi\right)-\partial_{\phi}V-\frac{1}{4}\partial_{\phi}\tau\, F^{2}-\frac{1}{2}\theta^{\prime}\,\sum_{i}(\partial \chi_i)^{2}=0\,,\nonumber\\
& \frac{1}{\sqrt{-g}}\partial_\mu\left(\sqrt{-g}\,\partial^\mu\psi\right)-\partial_{\psi}V-\frac{1}{4}\partial_{\psi}\tau\, F^{2}
-\frac{1}{2}\theta_{1}'\,\sum_{i}(\partial \sigma_i)^{2}=0\,,\nonumber\\
& \frac{1}{\sqrt{-g}}\partial_\mu\left(\theta_1\sqrt{-g}\,\partial^\mu\sigma_i\right)=0\,,\quad \frac{1}{\sqrt{-g}}\partial_\mu\left(\theta\sqrt{-g}\,\partial^\mu\chi_i\right)=0\,,\nonumber\\
& \partial_\mu(\sqrt{-g}\, \tau F^{\mu \nu})=0\,.
\end{align}
We now move on to discuss solutions of this theory. The above equations of motion admit a unit radius $AdS_4$ solution with vanishing matter fields, dual to the vacuum of a $d=3$ CFT with a conserved $U(1)$ charge. Placing the CFT at finite temperature and chemical potential corresponds to considering the Reissner-Nordstrom black hole in the bulk. However, as explained in section \ref{sec:setup} in this work we are interested in density wave states in the presence of a background lattice. Such states are described by the ansatz
\begin{align} \label{eq:DC_ansatz_num}
ds^{2}&=-U(r)\,dt^{2}+\frac{1}{U(r)}\,dr^{2}+e^{2V_{1}(r)}\,dx^1 dx^1 +e^{2V_{2}(r)}\,dx^2 dx^2\,,\nn
A&=a(r)\,dt\,,\nn
\phi&=\phi(r)\,,\qquad\qquad \chi_{i}=k_{i} x^{i}\,,\nn
\psi&=\psi(r)\,,\qquad\qquad \sigma_{i}=k_{si} x^{i}\,,
\end{align}
where $i=1,2$ (no summation). We now move on to specify boundary conditions.  In the IR, we demand the presence of a regular Killing horizon at $r=0$ by imposing the following expansion
\begin{align}\label{nhexpbh_num}
U\left(r\right)&=4\pi\,T\,r+\cdots\,,\qquad V_{i}=V_{i}^{(0)}+\cdots\,,\qquad a=a^{(0)}\,r+\cdots\,,\nn
\phi&=\phi^{(0)}(x)+\cdots\,,\qquad \psi=\psi^{(0)}(x)+\cdots\,,
\end{align}
which is specified in terms of 6 constants. In the UV, we demand the conformal boundary expansion
\begin{align}\label{asymptsol_num}
U&\to r^2+\cdots+W\,(r+R)^{-1}+\cdots,\qquad V_{1}\to \log(r+R)+\cdots+W_p (r+R)^{-3}+\cdots,\nn V_{2}&\to \log(r+R)+\cdots,\qquad\qquad\qquad a\to\mu+Q\,(r+R)^{-1}+\cdots,\nn
\phi&\to \phi_{s}\,(r+R)^{-1}+ \phi_{v}\,(r+R)^{-2}+\cdots, \qquad \psi\to  \psi_{s}\,+\cdots+\psi_{v}\,(r+R)^{-3}+\cdots\,.
\end{align}
Note, in particular, that we will take $\psi_s \neq 0$, and thus the scalar fields $(\psi,\sigma)$ constitute an anisotropic Q-lattice in which both translational invariance and $U(1)_\psi$ are explicitly broken. In the majority of this section, we will also demand that $\phi_s = 0$ in order for the density wave phase that is supported by $(\phi,\chi)$ to break the $U(1)_\phi$ spontaneously. Thus, this expansion is parametrized by 8 constants, as well as $k_{i},k_{si}$, which makes 12 constants in total. Overall we have 18 constants, in comparison to the 11 integration constants of the problem. Thus, for fixed $\gamma, \delta$ and temperatures below a critical one $T<T_c$, we expect to find a 7 parameter family of solutions, labelled by $k_{i},k_{si},\psi_{s},\mu,T$. These thermal states realise the scenario discussed in the previous sections and consequently, we expect all the results of section \ref{sec:pert} and \ref{sec:hydro} to apply. 

\begin{figure}[h!]
\centering
\includegraphics[width=0.48\linewidth]{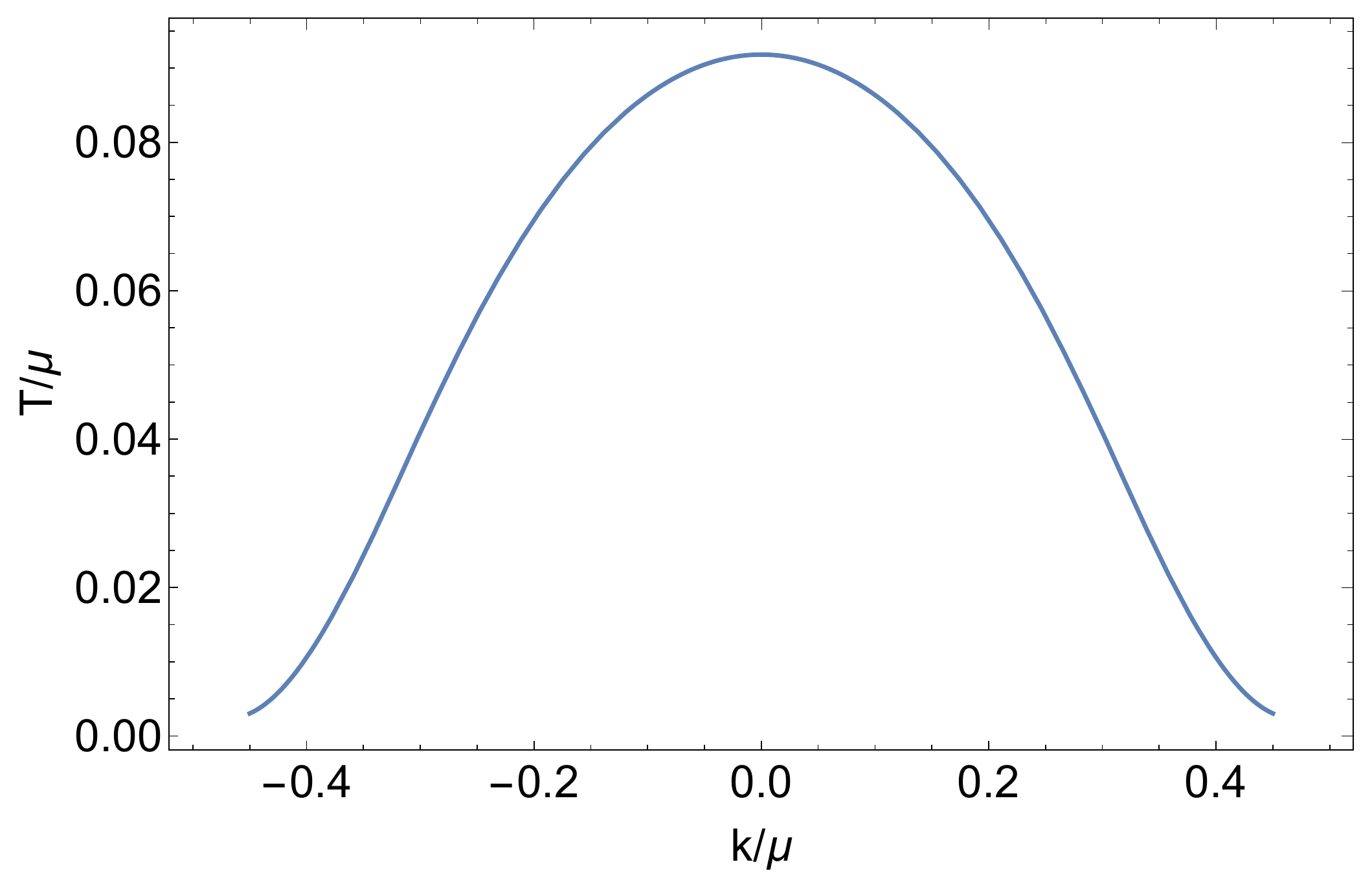}
\caption{Plot of the critical temperature at which the background Q-lattice becomes unstable as a function of $k/\mu$ for $(k_{s1}/\mu,k_{s2}/\mu,\psi_s,\gamma, \delta)=(0.3,0.3,4,3,1)$. We see that the most unstable mode corresponds to $k=0$ -- the thermodynamically dominant branch.}
\label{fig:BellCurve}
\end{figure}

In figure \ref{fig:BellCurve} we plot the critical temperature, $T_c$, as a function of $k$ for a particular choice of parameters. This is obtained by considering linearised fluctuations around the normal phase of the system ($\phi=0,\chi=0$) and exhibits the usual `Bell Curve'  shape.

\subsection{Quasinormal modes}

We now move on compute the spatially resolved quasinormal modes for the subclass of isotropic backgrounds constructed in the previous subsection that are characterised by $k_1=k_2 \equiv k, k_{s1}=k_{s2} \equiv k_s, V_1=V_2$. We take perturbations of the form
\begin{align}
\label{eq:deltag}
&\delta ds^2= - U \delta h_{tt} dt^2+2U \delta h_{t\,x_1}dt dx_1+ \delta h_{x_i x_i}dx^i dx^i\,,
\end{align}
together with $(\delta a_t, \delta a_{1}, \delta\phi,\delta\psi,\delta\chi_1,\delta\sigma_1)$, where the variations
are taken to be functions of $(t,r,x_1)$. We Fourier decompose our perturbations as
\begin{equation}
f(t,r)=e^{-i \omega v(t,r)+i q x_1}f(r)\,,
\end{equation}
where $v$ is the Eddington-Finkelstein coordinate defined as
\begin{align}
v(t,r,x_1)=t+\int_{\infty}^{r}\frac{dy}{U(y)}\,.
\end{align}
Note that our choice for the momentum $q$ to point in the direction $x_1$ is without loss of generality, because the background is isotropic. Plugging this ansatz in the equations of motion, we obtain 4 first order ODEs and 6 second order giving rise to 16 integration constants.

We now outline the boundary conditions for the fields. In the IR, we impose infalling boundary conditions at the horizon, which is $r=0$,
\begin{align}
&\delta h_{tt}=c_1 r^2+\cdots\,,&\delta h_{t \,x_1}=c_2 r+\cdots\,,\nonumber\\
&\delta h_{x_1x_1}=-c_3+\cdots\,,&\delta h_{x_2 x_2}=c_3+\cdots\,,\nonumber\\
&\delta a_t =c_4 r+\cdots\,,&\delta a_{1}=c_5+\cdots\,,\nonumber\\
&\delta\phi=c_6+\cdots\,,&\delta\psi=c_7+\cdots\,,\nonumber\\
&\delta\chi_1=c_8+\cdots\,,&\delta\sigma_1=c_9+\cdots\,,
\end{align}
where the constants $c_1,c_2$ and $c_4$ are not free but are fixed in terms of the others. Thus, for fixed value of $q$, we see that the expansion is fixed in terms of 7 constants, $\omega,c_3,c_5,c_6,c_7,c_8,c_9$. In the UV, the most general expansion one can write down is given by
\begin{align}
\label{eq:pertUV}
&\delta h_{tt}=\delta h_{tt}^{(s)}+\cdots\,,\qquad&\delta h_{t x_1}=\delta h_{t\,x_1}^{(s)}+\cdots\,,\nonumber\\
&\delta h_{x_1x_1}=\delta h_{x_1\,x_1}^{(s)}+\cdots\,,\qquad&\delta h_{x_2 x_2}=\delta h_{x_2\,x_2}^{(s)}+\cdots+\frac{\delta h_{22}^{(v)}}{(r+R)^3}+\cdots\,,\nonumber\\
&\delta a_t =a_{t}^{(s)}+\cdots\,,\qquad&\delta a_{1}=a_{1}^{(s)}+\frac{a_{1}^{(v)}}{(r+R)}+\cdots\,,\nonumber\\
&\delta\phi=\frac{\delta\phi^{(s)}}{(r+R)}+\frac{\delta\phi^{(v)}}{(r+R)^2}+\cdots\,,\qquad&\delta\psi=\delta\psi^{(s)}+\cdots+\frac{\delta\psi^{(v)}}{(r+R)^3}+\cdots\,,\nonumber\\
&\delta\chi_1=\delta\chi^{(v)}+\cdots\,,\qquad&\delta\sigma_1=\delta\sigma^{(s)}+\cdots+\frac{\delta\sigma^{(v)}}{(r+R)^3}+\cdots\,,
\end{align}
For the computation of quasinormal modes, we need to ensure that we remove all the sources from the UV expansion up to a combination of coordinate reparametrisations and gauge transformations
\begin{align}\label{eq:coord_transf}
[\delta g_{\mu\nu}+\mathcal{L}_\zeta g_{\mu\nu}]\to 0\,,\notag\\
[\delta A+\mathcal{L}_\zeta A+d\Lambda]\to 0\,.
\end{align} 
where the gauge transformations are of the form
\begin{align}\label{eq:coord_transf2}
x^\mu\to x^\mu+\zeta^\mu\,\quad\zeta&=e^{-i\omega t+i q x_1}\,\zeta^\mu \,\partial_\mu\,,\notag\\
A_\mu\to A_\mu+\partial_\mu \Lambda\,\quad\Lambda&=e^{-i\omega t+i q x_1}\,\lambda\,,
\end{align}
for $\zeta$, $\lambda$ constants. This requirement boils down to the sources apppearing in \eqref{eq:pertUV} taking the form
 \begin{align}
 \delta h_{t t}^{(s)}&=2 i \omega\, \zeta_1-2 \zeta_2 \notag\\
 \delta h_{t x_{1}}^{(s)}&=iq\, \zeta_1+i \omega \,\zeta_3\notag\\
 \delta h_{x_1\, x_{1}}^{(s)}&=-2 \zeta_2-2 i q \,\zeta_3\notag\\
 \delta h_{x_2\, x_{2}}^{(s)}&=-2 \zeta_2\notag\\
 \delta a_{t}^{(s)}&=i \mu\,\omega\, \zeta_1+i \omega \lambda\notag\\
 \delta a_{x_1}^{(s)}&=-i \mu\,q\, \zeta_1-i q\,\lambda\notag\\
 \delta\phi^{(s)}&=0\notag\\ 
\delta\psi^{(s)}&=0\notag\\ 
 \delta \sigma^{(s)}&=-k_{s}\,\zeta_3\,.
 \end{align}

We now see that the UV expansion is fixed in terms of 10 constants: $\zeta_1,\zeta_2,\zeta_3, \lambda$ and $\delta h_{x_2\,x_2}^{(v)},a_{1}^{(v)},\delta\phi^{(v)},\delta\psi^{(v)},\delta\sigma^{(v)},\delta\chi^{(v)}$. Overall, for fixed $q$, we have 17 undetermined constants, of which one can be set to unity because of the linearity of the equations. This matches precisely the 16 integration constants of the problem and thus we expect our solutions to be labelled by $q$.

\begin{figure}[h!]
\centering
\includegraphics[width=0.48\linewidth]{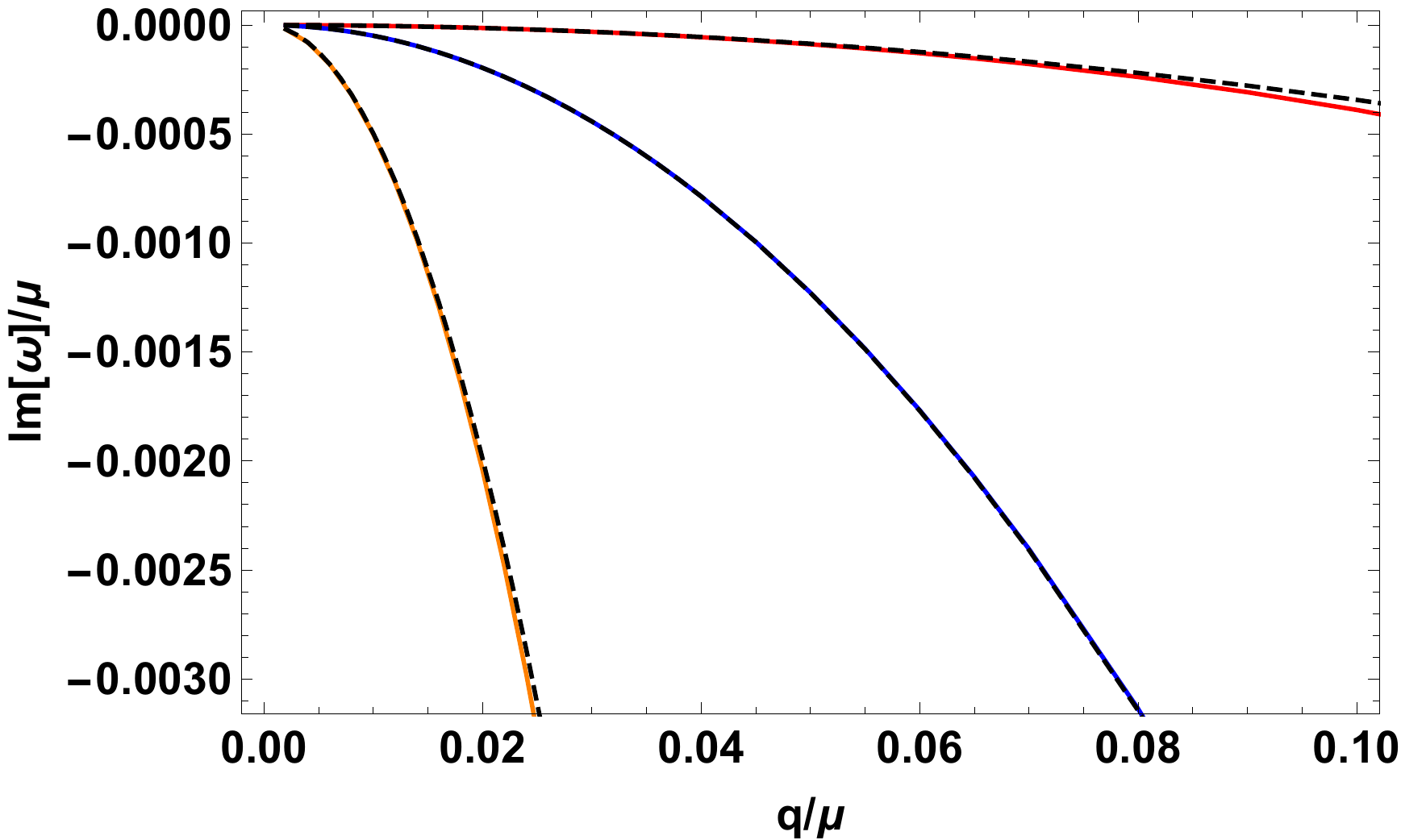}\quad\includegraphics[width=0.48\linewidth]{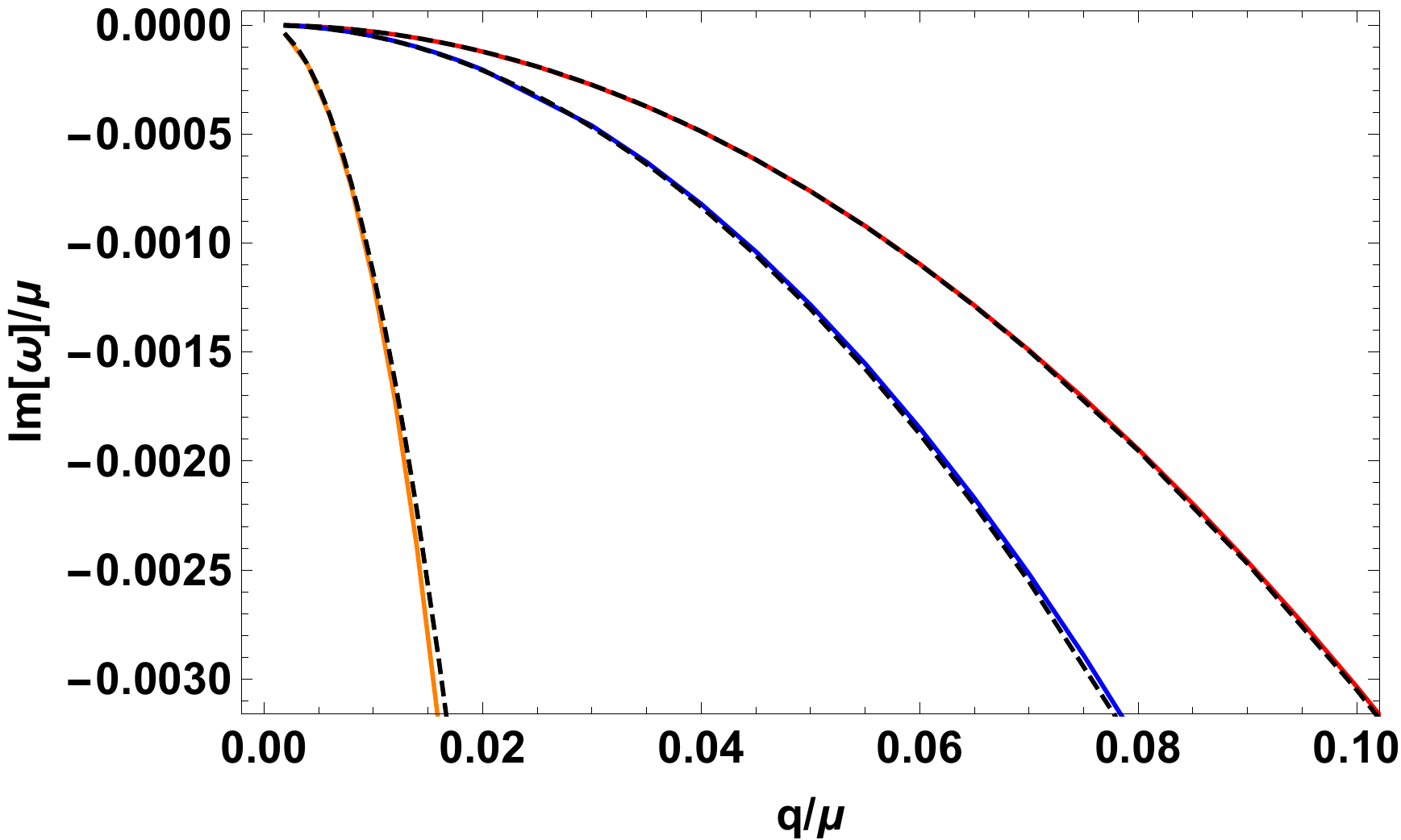}\\
\caption{The dispersion relation for our three diffusive modes. Both panels are at $(k_s/\mu,T/\mu,\psi_s,\gamma, \delta)=(0.3,0.04,4,3,1)$, whilst the left panel is at $k/\mu=0$ and the right panel is at $k/\mu = 0.15$. The dashed lines represent the dispersion relations obtained from the linear system \eqref{3by3}. In the $k\to 0$ limit the blue modes become the $U(1)$ Goldstone \cite{Donos:2019txg}, and the red modes become the incoherent mode described in \cite{Davison:2015taa}.}
\label{fig:modes}
\end{figure}

We proceed to solve numerically this system of equations subject to the above boundary conditions using a double-sided shooting method. We find our ansatz contains three hydrodynamic modes as expected, with diffusion constants in quantitative agreement with the analytical predictions \eqref{xmatrix} and \eqref{sigmamatrix}. Figure \ref{fig:modes} displays the dispersion relations of our QNMs at $k=0$ and at a moderately higher value of $k$. In the same figure, we also illustrate the good agreement between the numerical computation and the dispersion relations fixed by the linear system \eqref{3by3} of our analytic treatment, in the $q\to 0$ limit.

\begin{figure}[h!]
\centering
\includegraphics[width=0.48\linewidth]{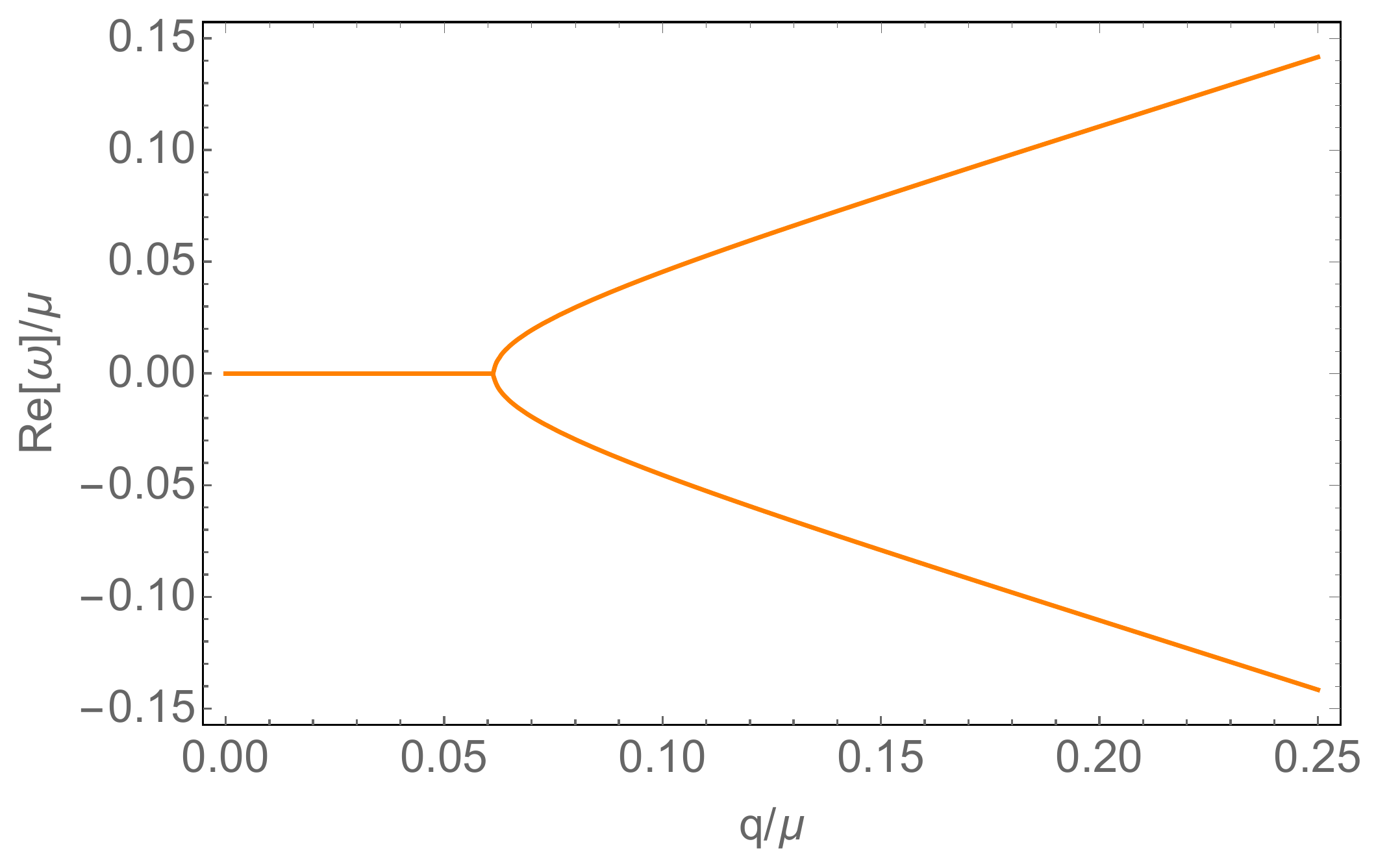}\quad\includegraphics[width=0.48\linewidth]{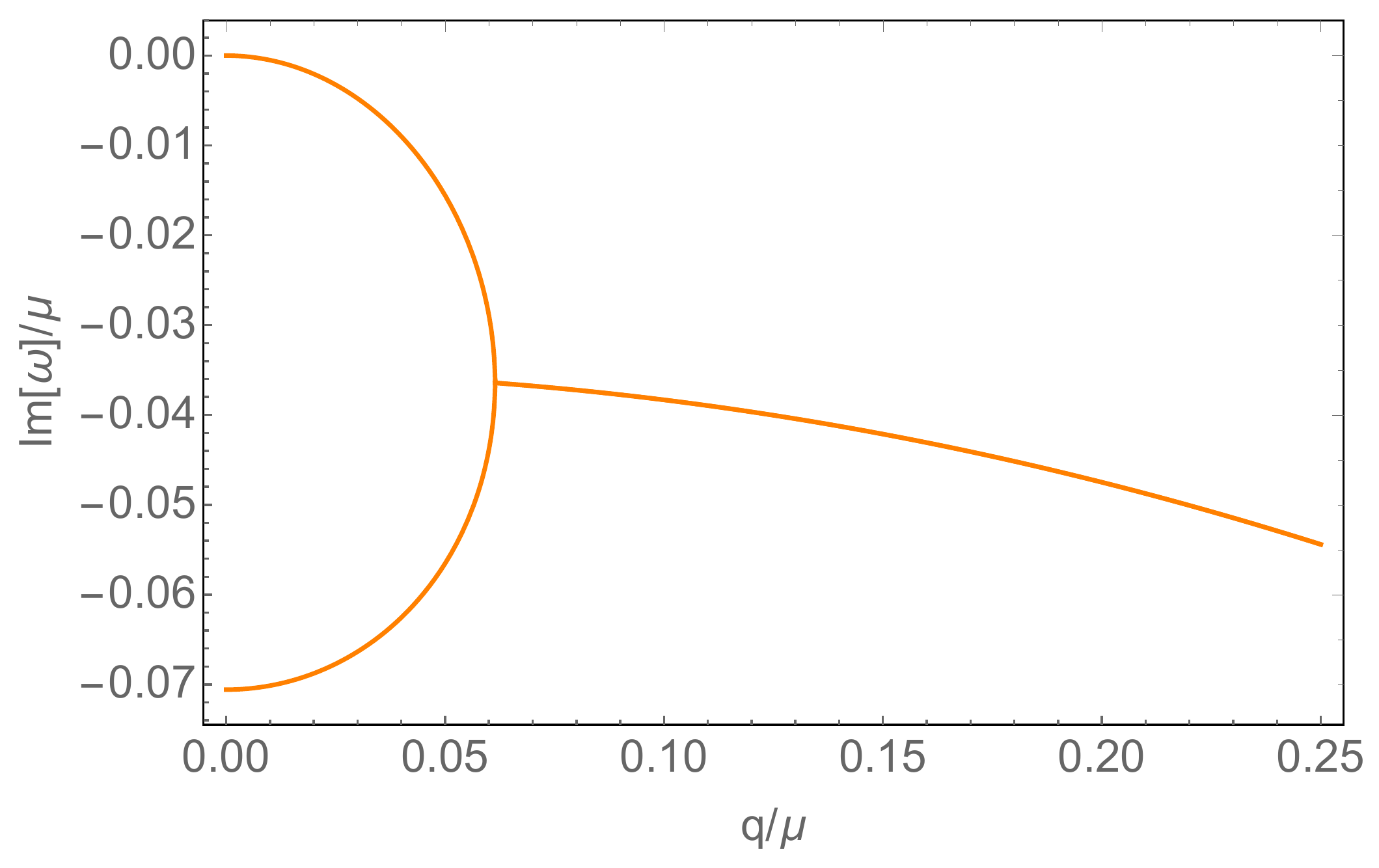}\\
\caption{Plots of Re[$\omega$]$/\mu$ and Im[$\omega$]$/\mu$ as functions of $q/\mu$ for $(k_s/\mu,T/\mu,\psi_s,\gamma, \delta)=(0.3,0.04,4,3,1)$ and $k/\mu=0.15$, showing the strongest of our diffusive modes and the momentum relaxation mode undergoing a collision to form two sound modes as $q$ is raised from zero.}
\label{fig:collision}
\end{figure}

Figure \ref{fig:collision} shows how, as $q$ is raised, one of the modes collides with the momentum relaxation mode to form two modes which behave like sound modes, as expected by the hydrodynamic crossover. As $k_s \rightarrow 0$, the $q=0$ momentum mode is lowered until at $k_s = 0$ it and the diffusive mode disappear completely, leaving only sound modes, the mixture of the Goldstone mode and the incoherent mode of \cite{Davison:2015taa}.

\subsection{AC conductivities and the gap}

\begin{figure}[h!]
\centering
\includegraphics[width=0.48\linewidth]{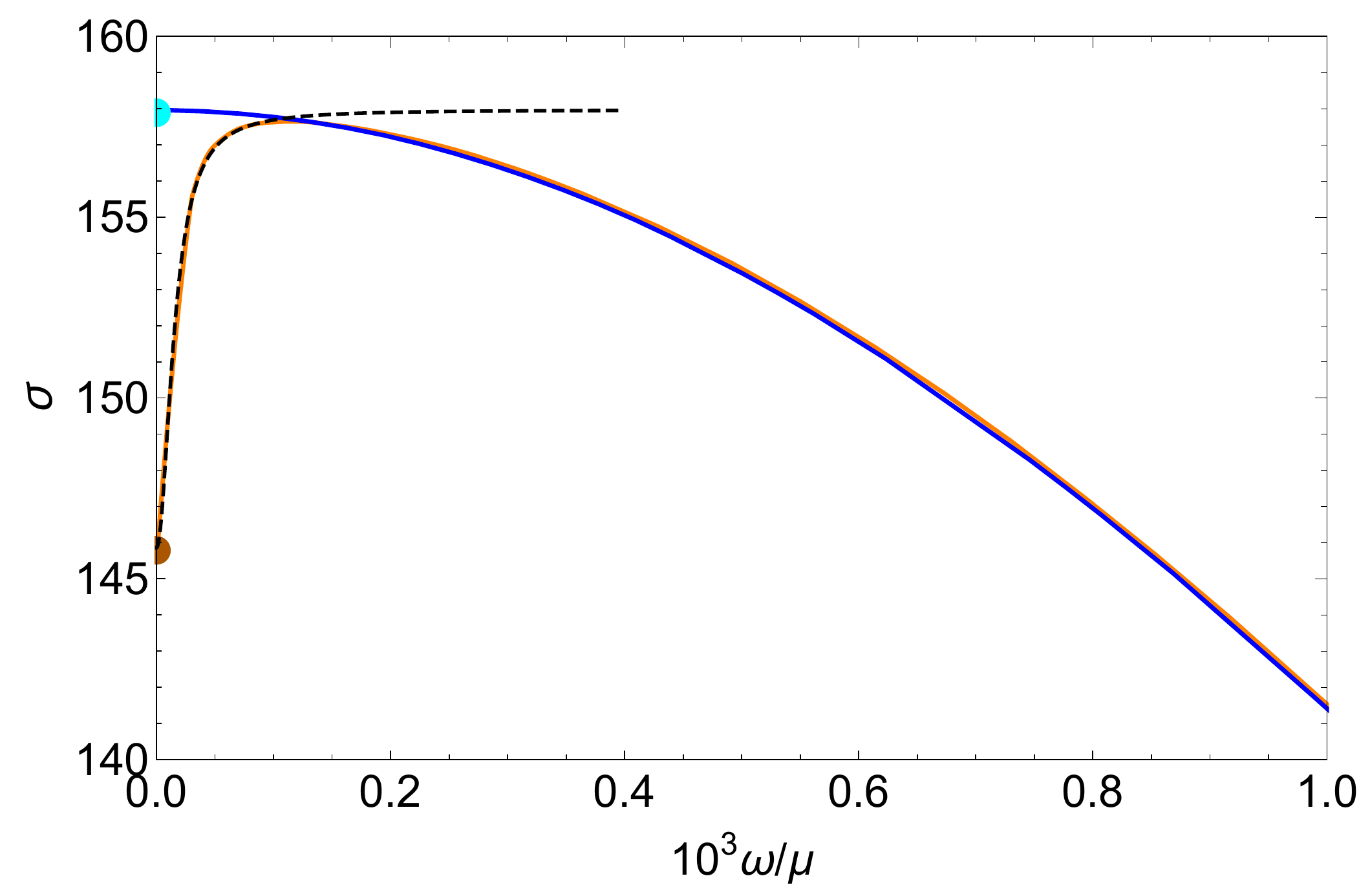} \includegraphics[width=0.48\linewidth]{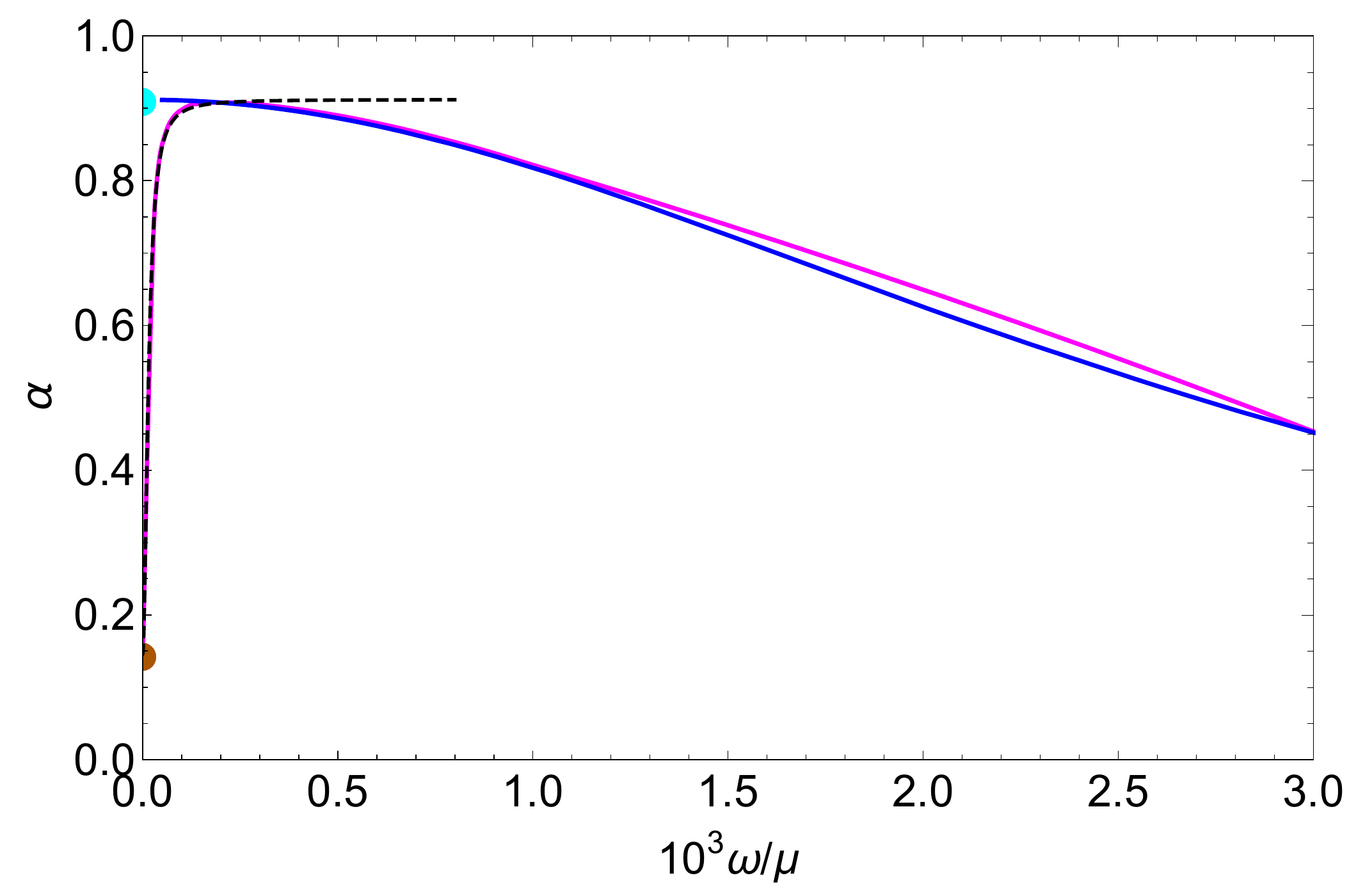}\\
\includegraphics[width=0.48\linewidth]{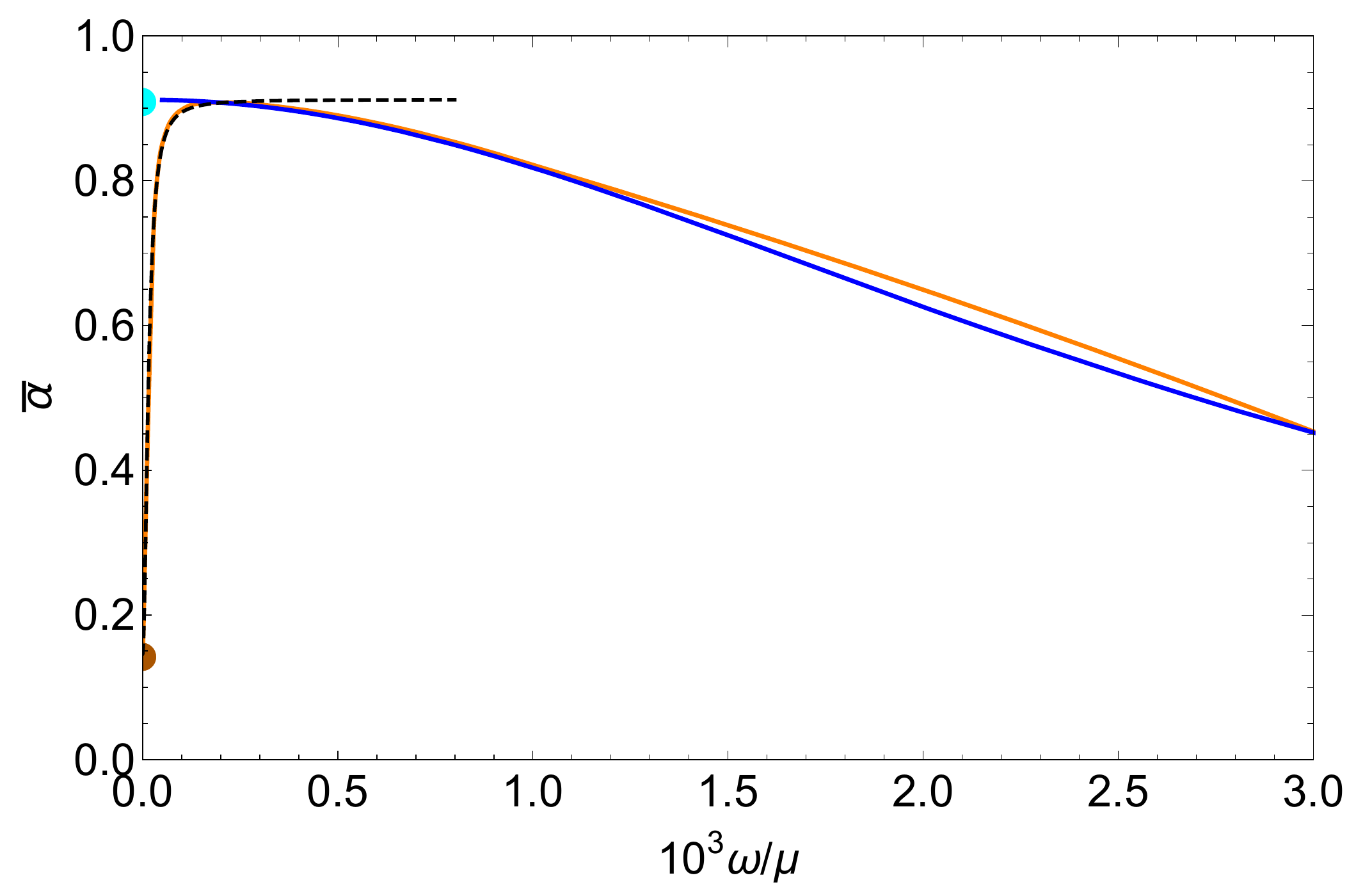} \includegraphics[width=0.48\linewidth]{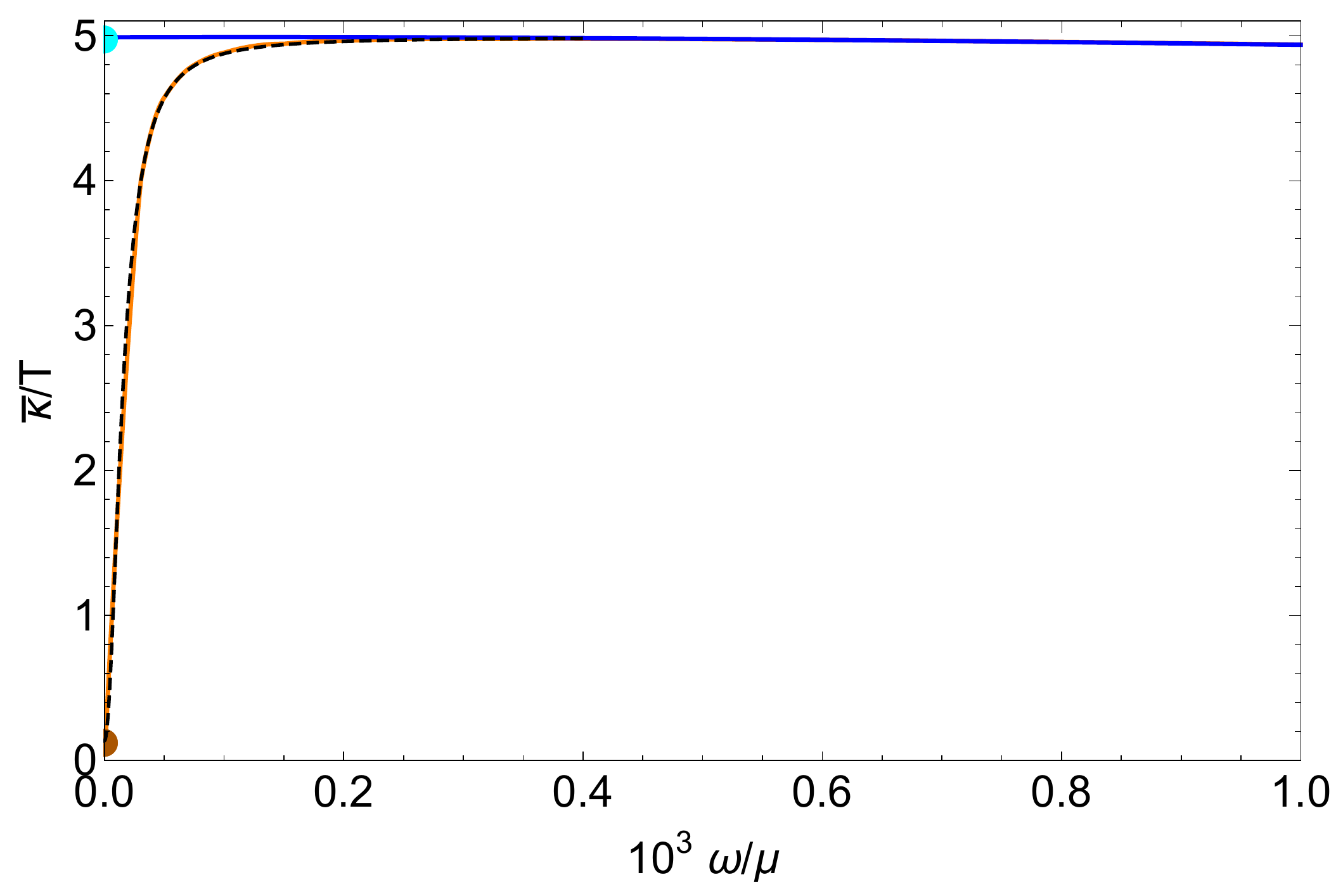}
\caption{Plot of the electric conductivity as a function of the frequency for $\{k/\mu,k_s/\mu,T/\mu,\psi_s,\gamma, \delta\}=\{0.15,0.3,0.01,4,3,0.5\}$ and for $\phi_{s}=0$ (blue) and $\phi_{s}/\mu=10^{-5}$ (orange). The black, dashed line corresponds to the analytic result \eqref{eq:AC_cond}. The orange line corresponds to the numerical result for the conductivities as functions of the frequency for $\phi_{s}/\mu=10^{-5}$, which yelds $\omega_g=1.54 \cdot 10^{-5}$, while the blue line corresponds to the case without pinning, $\phi_{s}=0$.}
\label{fig:AC_cond_pinning}
\end{figure}

We begin by comparing the analytic results for the AC thermoelectric conductivities \eqref{eq:AC_cond} with the full numerical calculation carried out in section (5.2) of \cite{Donos:2019tmo} in the presence of pinning in the model \eqref{eq:bulk_action3},\eqref{model}. The results are shown in Figure \ref{fig:AC_cond_pinning}, for $\{k/\mu,k_s/\mu,T/\mu,\psi_s,\gamma, \delta\}=\{0.15,0.3,0.01,4,3,0.5\}$.  We see very good quantitative agreement for frequencies $\omega\le\omega_g$.

\begin{figure}[h!]
	\centering
	\includegraphics[width=0.48\linewidth]{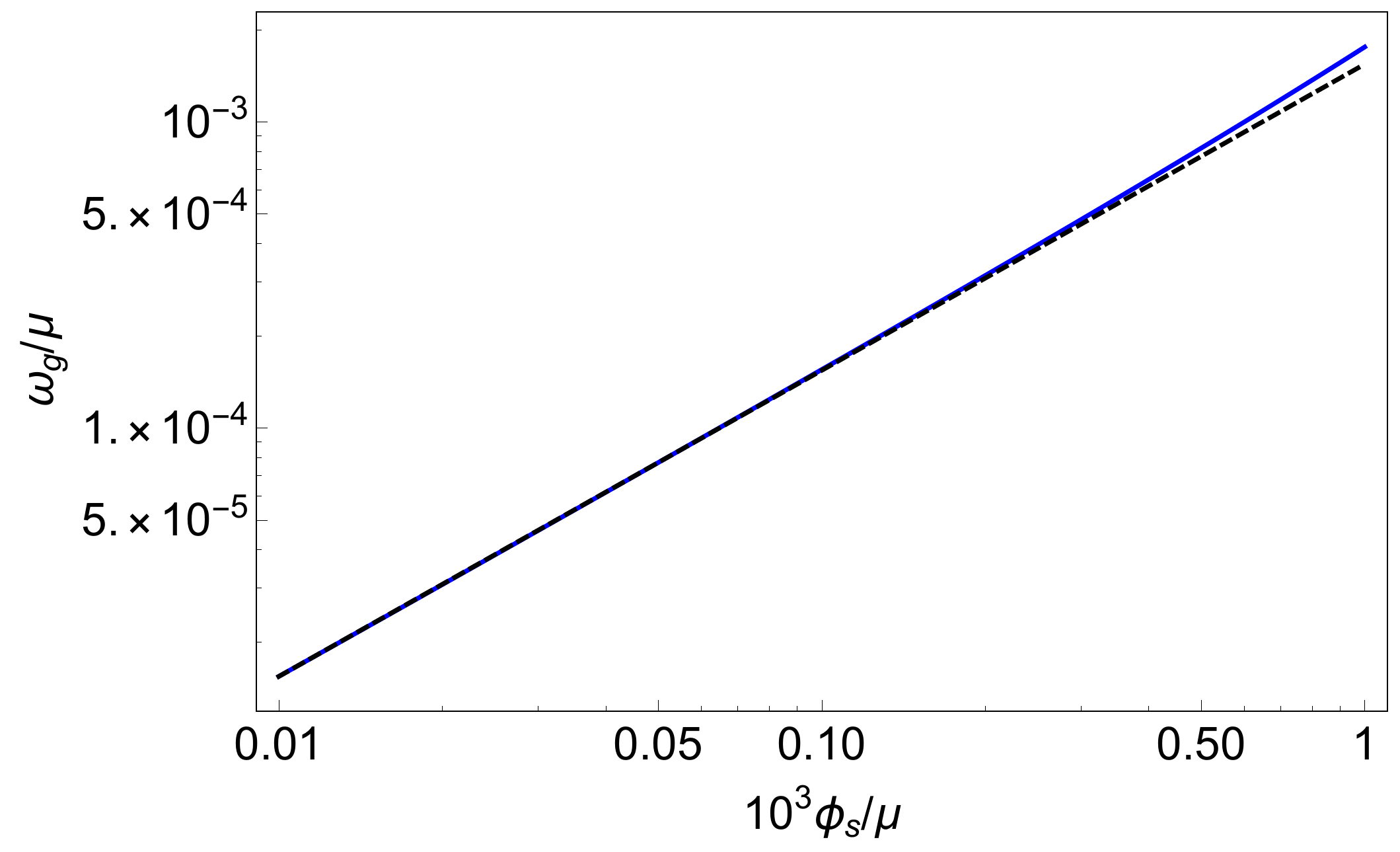}
	\caption{Plot of the gap as a function of the pinning parameter $\phi_s$ for $\{k/\mu,k_s/\mu,T/\mu,\psi_s,\gamma, \delta\}=\{0.15,0.3,0.01,4,3,0.5\}$. The solid blue line corresponds to the numerical computation for the gap, while the black dashed line corresponds to \eqref{eq:gap} evaluated on the same background black hole.}
	\label{fig:gap_pinning}
\end{figure}

Let us also compare the analytic formula for the gap to a full numerical calculation in the model \eqref{eq:bulk_action3}\eqref{model}. In the numerics, we used the same set up and expansions as in section (5.2) of \cite{Donos:2019tmo} in the presence of pinning, but we set the external sources to zero. In turn, this constrains the value of the frequency such that a non-trivial solution could be found. The results are shown in Figure \ref{fig:gap_pinning}, again for $\{k/\mu,k_s/\mu,T/\mu,\psi_s,\gamma, \delta\}=\{0.15,0.3,0.01,4,3,0.5\}$. We see that there is good quantitative agreement for small pinning parameter $\phi_s$, confirming our analytic computation. This extends the results of \cite{Donos:2019txg} to include the mixing of the bulk Goldstone to the heat current.

\section{Discussion}\label{sec:discussion}

In this paper we considered thermal phases of holographic lattices at finite chemical potential which exhibit spontaneous breaking of a global symmetry in the bulk. Even though such a symmetry breaking in the bulk doesn't imply the breaking of a continuous symmetry on the boundary, we expect the emergence of a diffusive mode from the field theory point of view.

We took the order parameter to break translations itself, resulting to the coupling of the corresponding hydrodynamic sliding mode to the heat and electric currents of the theory. In the unbroken phase and with the translations being broken explicitly by the lattice, the long wavelength excitations of the conserved charges are well described by incoherent hydrodynamics. In that regime, late time dynamics at long distances is dominated by two thermoelectric diffusive modes. In section \ref{sec:hydro} we incorporated the emergent sliding mode in order to give an hydrodynamic description which is valid below the critical temeprature.

Our results clarify the role of the horizon thermoelectric coefficients as transport coefficients appearing in our derivative expansion. We further considered the driven hydrodynamics of our system by including a time dependent external electric field and temperature gradient. More interestingly, in our description we took into account the effect of a perturbative static source which adds a small pinning parameter for our spontaneous density wave. This allowed us to compute the optical conductivities of our system and give explicit formulae demonstrating the transfer of spectral weight to frequencies set by the gap of the theory.

This explained previous numerical results in the literature \cite{Donos:2019tmo} for models that realise the same mechanism that we are proposing here. One could argue that the theory of hydrodynamics presented in sections \ref{sec:hydro} and \ref{sec:pinning} is also applicable to systems without a gravitational dual at frequency scales much lower than the momentum relaxation rate. In building such a theory, one would have to write a general ``Josephson relation'' \eqref{eq:dy_source} which couples the phase $\delta\hat{c}_{g}$ to the currents of theory. In the lack of an almost conserved momentum operator, these currents can only be the universal heat and electric currents. It is important to notice that for our holographic model, it is only the heat current that appears in our equation \eqref{eq:dy_source}, even in the presence of finite chemical potential. We believe that this is an artefact of the simplicity of our model \eqref{eq:bulk_action}.

A direct comparison of our results with previous holographic computations \cite{Andrade:2017ghg,Andrade:2017leb,Andrade:2017cnc,Andrade:2015iyf,Amoretti:2018tzw,Andrade:2018gqk,Amoretti:2019cef} is not clear as in those setups the momentum and phase relation rate were governed by the same scale. However, we should be able to compare with \cite{Delacretaz:2017zxd} where the momentum relaxation rate $\Gamma$ was taken to be parametrically small but independent of the phase relaxation rate $\Omega$, in their notation. The comparison needs to be done in the limit $\omega\sim \Omega\ll\Gamma$ and with $\Gamma$ smaller than any other scale such as the temperature and the chemical potential. In this limit, we can obtain a version of incoherent hydrodymamics by integrating out the fluid velocity from the theory of \cite{Delacretaz:2017zxd} which we present in Appendix \ref{app:weak_mom_relax}. We compare and contrast to a limit of our theory in which the strength of the explicit lattice is parametrically small as compared to all the other scales and much greater than the pinning parameter $\delta\phi_{1s}$.

It is worth examining the behaviour of the gap \eqref{eq:gap} as we approach the critical temperature $T\to T_{c}$ where the would be gapless mode makes its appearance. Our approximations remain valid as long as $(1-T/T_{c})^{1/2}\gg\varepsilon$. In that case, we have that all the quantities that appear in \eqref{eq:gap} remain finite apart from $\Phi^{(0)}\sim 1-T/T_{c}$ and $ \left | \langle \mathcal{O}_{Z_{1}}\rangle \right |\sim (1-T/T_{c})^{1/2}$ suggesting that $\omega_{g}\sim (1-T/T_{c})^{-1/2}\,\delta\phi_{1s}$. Close to the critical temperature, this in a sense similar to the empirical observation made in \cite{Ammon:2019wci}. One can imagine that with specific choices of the functions that appear the action \eqref{eq:bulk_action}, the ground states of our theory will be similar to those described in e.g. \cite{Donos:2014uba,Gouteraux:2014hca}. For those ground states, our formula \eqref{eq:gap} will be powerful enough to predict the behaviour of the gap away from the critical regime, at low temperatures.

We have greatly benefited from the simplicity of our Q-lattice construction. One might wonder whether our results can be naturally extended to more general holographic models which don't require a global symmetry in the bulk and which involve inhomogeneous black holes. This is a natural question to ask and the techniques developed in \cite{Donos:2017gej} and here should help in making progress towards this direction.

\section*{Acknowledgements}
 
We would like to thank B. Gout\'{e}raux for helpful correspondence. VZ would like to thank Matteo Baggioli for interesting discussions. AD, DM and CP are supported by STFC grant ST/P000371/1. VZ is supported by the China Postdoctoral Science Foundation (International Postdoctoral Fellowship Program 2018) and the National Natural Science Foundation of China (NSFC) (Grant number 11874259).
%\newpage

\appendix

\section{Derivative expansion of bulk perturbations}\label{app:hydro_expansion}

We begin the bulk construction of the hydrodynamic modes of the model \eqref{eq:bulk_action} by analysing their long wavelength limit, $q \rightarrow 0$, $\omega \rightarrow 0$. We wish to work in a derivative expansion around this seed. We begin by giving more details on the background thermodynamic perturbations $\delta X_b$ which were introduced in the main text in equation \eqref{eq:deltaX0}.

If we are to work in Schwarzschild co-ordinates, a naive variation $T \rightarrow T + \delta T_{[0]}$, $\mu \rightarrow \mu + \delta \mu_{[0]}$ of our background will take us outside the class of zero-source, infalling solutions, as explained in \cite{Donos:2017ihe}. However, our solution is related to genuine quasinormal modes by co-ordinate and gauge transformations. In particular, we can perform the coordinate transformation $t\to t + \frac{\delta T_{[0]}}{T}\,g(r)$ in which $g(r)$ vanishes sufficiently fast as $r\to \infty$ and $g(r)\to \ln r/(4\pi T)+g^{(1)}r+\cdots$ as $r\to 0$, along with the gauge transformation $\delta A=d\Lambda$ with $\Lambda=-\left( t+g(r)\right)\delta\mu_{[0]}$.

At the horizon, this leaves us with the $r$-expansion
\begin{align}\label{eq:nh_static_metric_pertb}
\delta ds^{2}_{[0]}&=-\frac{\delta T_{[0]}}{T}\,\left(4\pi T r\,dt^{2}+\frac{dr^{2}}{4\pi T r} \right) -2\,\frac{\delta T_{[0]}}{T}\,dt\,dr +\left(\frac{\partial e^{2V_{1}^{(0)}}}{\partial T}\,\delta T_{[0]}+\frac{\partial e^{2V_{1}^{(0)}}}{\partial\mu}\,\delta\mu_{[0]}\right)dx^{1} dx^{1}\nn
&+\left(\frac{\partial e^{2V_{2}^{(0)}}}{\partial T}\,\delta T_{[0]}+\frac{\partial e^{2V_{2}^{(0)}}}{\partial\mu}\,\delta\mu_{[0]}\right)dx^{2} dx^{2} +\cdots\,,\nn
\delta a_{[0] t}&=-\delta\mu_{[0]} +r\,\left(\frac{\partial a_{t}^{(0)}}{\partial T}\,\delta T_{[0]}+\frac{\partial a_{t}^{(0)}}{\partial\mu}\,\delta\mu_{[0]}\right)+\cdots\,,\qquad \delta\chi_{1[0]}=\delta c_{g[0]}\,,\nn
\delta a_{[0]r}&=-\delta\mu_{[0]}\,(4\pi T\,r)^{-1}+\frac{\delta T_{[0]}}{T}(4\pi T)^{-1}a_{t}^{(0)} -g^{(1)}\delta\mu_{[0]}+\cdots\,,\nn
\delta\phi_{[0]}&=\frac{\partial \phi^{(0)}}{\partial T}\,\delta T_{[0]}+\frac{\partial \phi^{(0)}}{\partial\mu}\,\delta\mu_{[0]}+\cdots\,,\qquad \delta\psi_{[0]}=\frac{\partial \psi^{(0)}}{\partial T}\,\delta T_{[0]}+\frac{\partial \psi^{(0)}}{\partial\mu}\,\delta\mu_{[0]}+\cdots\,.
\end{align}
The above is exactly the near horizon limit of the perturbation $\delta X_{[0]}$ as defined in \eqref{eq:deltaX0}.

Turning on $\varepsilon$, and continuing to demand regularity at the horizon, we see that the behaviour of the corresponding hydrodynamic modes at the horizon is governed by
\begin{align}\label{eq:hor_fluid_exp}
\omega&=\varepsilon\,\omega_{[1]}+\varepsilon^{2}\,\omega_{[2]}+\cdots\,,\nn
p&=4\pi \left( \delta T_{[0]}+\varepsilon\, \delta T_{[1]}+\varepsilon^2\, \delta T_{[2]}  +\cdots\right)\,,\nn
v&=\varepsilon\,v_{[1]}+\varepsilon^2\,v_{[2]}\cdots\,,\nn
\varpi&=\,- \left( \delta\mu_{[0]} + \varepsilon\,\delta\mu_{[1]} +\varepsilon^2\,\delta\mu_{[2]}+\cdots\,\right),\nn
\delta g_{ij}^{(0)}&=\,\left(\frac{\partial g^{(0)}_{ij}}{\partial T}\,\delta T_{[0]}+\frac{\partial g^{(0)}_{ij}}{\partial \mu}\,\delta \mu_{[0]} +\varepsilon\, \delta g_{[1]ij}^{(0)}+\cdots\right)\,,\nn
\delta\phi^{(0)}_I&=\,\left(\frac{\partial \phi_I^{(0)}}{\partial T}\,\delta T_{[0]}+\frac{\partial \phi_I^{(0)}}{\partial\mu}\,\delta\mu_{[0]} +\varepsilon\,\delta\phi_{I[1]}^{(0)}+\cdots \right)\,,\nn
\delta\psi^{(0)}_I&=\,\left(\frac{\partial \psi_I^{(0)}}{\partial T}\,\delta T_{[0]}+\frac{\partial \psi_I^{(0)}}{\partial\mu}\,\delta\mu_{[0]} +\varepsilon\,\delta\psi_{I[1]}^{(0)}+\cdots \right)\,,\nn
\delta\chi^{(0)}_I&=\, \delta c_{g[0]}   \delta_{1I} +\varepsilon\,\delta\chi_{I[1]}^{(0)}+\cdots \,,\nn
\delta\sigma^{(0)}_I&=\,\varepsilon\,\delta\sigma_{I[1]}^{(0)}+\cdots \,,
\end{align}
where we used the definitions in \eqref{eq:nh_reg} for $p$, $v$ and $\varpi$.

\subsection{The equations at leading order in $\varepsilon$}\label{sec:e1}

We now begin the task of using the constraints \eqref{hconstrainta}-\eqref{hiconstraint2b}, and the eom for $\chi_1$ \eqref{eq:bulk_eom_chi}, to constrain the form of our hydrodynamic modes.
At $\mathcal{O}(\varepsilon)$, the vector constraint gives
\begin{align}\label{eq:vec1o1}
e^{V_{1}^{(0)}+V_{2}^{(0)}} \, \mathcal{B} \, v_{[1]} + iq \left(\rho\,\delta\mu_{[0]} +s\,\delta T_{[0]} \right) -i\omega_{[1]}\,k_{1} \, e^{V_{1}^{(0)}+V_{2}^{(0)}} \, \Phi_1^{(0)}\,\delta c_{g[0]} &=0\,,
\end{align}
in which we use the notation $\mathcal{B}$ from \eqref{eq:horizon_coeffs}. We will shortly be using this equation to express the horizon fluid velocity $v_{[1]}$ in terms of the parameters of the zero mode, $\delta c_{g[0]}$, $\delta T_{[0]}$ and $\delta \mu_{[0]}$.

Next, let us examine the $\chi_1$ eom at $\mathcal{O}(\varepsilon)$. We find
\begin{align}
&\partial_{r}\left[e^{V_{1}+V_{2}}\, \Phi_1 \left(-i\omega_{[1]} \delta c_{g[0]} + U\partial_{r}\delta \chi_{1[1]} - k_{1}\,e^{-2V_{1}}\, \delta g_{r1[1]}\right)\right]\nn
&+ iq \, e^{V_{2}-V_{1}}\, k_{1}\, \left( \partial_{\phi_I} \Phi_1 \, \delta \phi_{I[0]} + \partial_{\psi_I} \Phi_1 \, \delta\psi_{I[0]} + \frac{\Phi_1}{2} \left(- e^{-2V_{1}}\delta g_{11[0]} + e^{-2V_{2}}\delta g_{22[0]} \right) \right) = 0.
\end{align}
From this equation, we find asymptotic behaviour
\begin{align}
\delta\chi_{1[1]}&=\frac{r^{2\Delta_{Z_{1}}-3}}{(2\Delta_{Z_{1}}-3)\phi_{1v}^{2}}\,\left\{-e^{V^{(0)}_{1}+V^{(0)}_{2}} \Phi^{(0)}_1 \left( i\omega_{[1]} \delta c_{g[0]} - k_{1}v_{[1]}\right) \right.\nn
&\left. -iq\,k_{1}\int dr\,e^{V_2-V_1}\bigg[\partial_{\phi_I} \Phi_1 \, \delta \phi_{I[0]} + \partial_{\psi_I} \Phi_1 \, \delta\psi_{I[0]}+\Phi_1 \left(\delta V_{2[0]}-\delta V_{1[0]}\right) \bigg] \right\}+\cdots\,.\notag
\end{align}
Following our discussion on the asymptotics \eqref{eq:gen_UVexp} and by demanding that $\zeta_{S_{1}}=0$ we obtain the equation
\begin{align}\label{eq:intchi1o1}
e^{V^{(0)}_{1}+V^{(0)}_{2}} \Phi^{(0)}_1 \left(i\omega_{[1]} \delta c_{g[0]} - k_{1}v_{[1]}\right) = -iq\,k_{1}\int dr\,e^{V_2-V_1}\bigg[&\partial_{\phi_I} \Phi_1 \, \delta \phi_{I[0]} + \partial_{\psi_I} \Phi_1 \, \delta\psi_{I[0]}\nn
&+\Phi_1 \left(\delta V_{2[0]}-\delta V_{1[0]}\right) \bigg]\,.
\end{align}
Using \eqref{eq:kvariation}, we can express the RHS of \eqref{eq:intchi1o1} as $-iq$ times
\begin{align}
\delta T_{[0]}\frac{\delta}{\delta T} \left(\frac{\delta w_{FE}}{\delta k_{1}}\right) + \delta \mu_{[0]} \frac{\delta }{\delta \mu} \left(\frac{\delta w_{FE}}{\delta k_{1}}\right) = -\frac{\delta s}{\delta k_{1}} \delta T_{[0]} - \frac{\delta \rho}{\delta k_{1}} \delta\mu_{[0]} \,.
\end{align}
After substituting for $v_{[1]}$ from \eqref{eq:vec1o1}, the $\chi_1$ equation of motion then gives a relation between the parameters of the zero mode,
\begin{align}\label{eq:chi1o1final}
iq \left( k_1 \Phi_1^{(0)} s - \mathcal{B} \frac{\delta s}{\delta k_{1}} \right) \delta T_{[0]} &+ iq  \left( k_1 \Phi_1^{(0)} \rho - \mathcal{B} \frac{\delta \rho}{\delta k_{1}} \right) \delta \mu_{[0]} \nn
&+ i \omega_{[1]} \left( k_{s1}^2 \, \Phi^{(0)}_1 \Psi_1^{(0)} \, e^{V^{(0)}_{1}+V^{(0)}_{2}} \right) \delta c_{g[0]} = 0 \,.
\end{align}

However, this is not the only relation the horizon constraints give us. The scalar constraints at the horizon \eqref{hconstrainta}-\eqref{hconstrainttwo} read, at $\mathcal{O}(\varepsilon)$,
\begin{align}\label{eq:first_conflicting_relation}
&0=i\omega_{[1]}\left(\delta T_{[0]} \frac{\partial V^{(0)}_{1[0]}}{\partial T} +\delta\mu_{[0]} \frac{\partial V^{(0)}_{2[0]}}{\partial\mu}\right)\,,\nn
&0 =i\omega_{[1]} \,\bigg[\tau^{(0)}a^{(0)} \left( \frac{\partial V^{(0)}_{1[0]}}{\partial T} + \frac{\partial V^{(0)}_{2[0]}}{\partial T}\right)\nn
&\qquad+\partial_{\phi_I} \tau^{(0)}a^{(0)}\, \frac{\partial \phi_I^{(0)}}{\partial T}+\partial_{\psi_I} \tau^{(0)}a^{(0)}\, \frac{\partial \psi_I^{(0)}}{\partial T}+\tau^{(0)} \frac{\partial a_{t}^{(0)}}{\partial T}\bigg]\delta T_{[0]}\nn
&\quad+i\omega_{[1]} \,\bigg[\tau^{(0)}a^{(0)} \left( \frac{\partial V^{(0)}_{1[0]}}{\partial \mu} + \frac{\partial V^{(0)}_{2[0]}}{\partial \mu}\right)\nn
&\qquad+\partial_{\phi_I} \tau^{(0)}a^{(0)}\, \frac{\partial \phi_I^{(0)}}{\partial \mu}+\partial_{\psi_I} \tau^{(0)}a^{(0)}\, \frac{\partial \psi_I^{(0)}}{\partial \mu}+\tau^{(0)} \frac{\partial a_{t}^{(0)}}{\partial \mu}\bigg]\delta \mu_{[0]} \,.
\end{align}
As was demonstrated in \cite{Donos:2017ihe}, this system can be expressed in terms of thermoelectric susceptibilities,
\begin{align}\label{dcmat}
i\omega_{[1]}\left( \begin{array}{cc}
T^{-1}c_{\mu} &\xi\\
\xi &\chi_q
\end{array} \right)
\left( \begin{array}{c}
\delta T_{[0]}\\
\delta\mu_{[0]}
\end{array} \right)=0\,.
\end{align}
We now consider the two possibilities for $\omega_{[1]}$:

\textbf{$\omega_{[1]} \neq 0$:} Provided the matrix of susceptibilities in \eqref{dcmat} is invertible, as is generically the case, we deduce from \eqref{dcmat} that $\delta T_{[0]} = \delta\mu_{[0]} = 0$. However, when $k_{s1} \neq 0$ and momentum is relaxing, this means \eqref{eq:chi1o1final} is no longer solvable, excepting the trivial perturbation $\delta c_{g[0]} = 0$.

\textbf{$\omega_{[1]} = 0$:} In this case, \eqref{dcmat} contributes nothing new. However, at next order in $\varepsilon$, the scalar constraints will lead to another version of \eqref{dcmat}, but with $\omega_{[1]} \rightarrow \omega_{[2]}$. At $k_1 \neq 0$, the only way this relation can avoid conflicting with \eqref{eq:chi1o1final} is if $\delta T_{[0]} = \delta\mu_{[0]} = 0$.

To summarise, in the presence of momentum relaxation the hydrodynamic modes generated by our thermodynamic perturbations are diffusive, and in the presence of spontaneous breaking of the bulk global symmetry they are each seeded by a zero mode with $\delta c_{g[0]} \neq 0$, $\delta T_{[0]}=\delta\mu_{[0]}=0$.

\subsection{The equations at next-to-leading order in $\varepsilon$}

We proceed to constrain all three diffusion constants by use of the horizon vector constraint at $\mathcal{O}(\varepsilon^2)$, the equation of motion for $\chi_1$ at $\mathcal{O}(\varepsilon^2)$, and the horizon scalar constraints at order $\mathcal{O}(\varepsilon^3)$.

Expanding the vector constraint \eqref{hiconstraint2b} at $\mathcal{O}(\varepsilon^2)$ we obtain
\begin{align}
&e^{V^{(0)}_{1}+V^{(0)}_{2}}\mathcal{B} v_{[2]} + iq \left(s\delta T_{[1]} + \rho\delta\mu_{[1]}\right) -i\omega_{[2]} k_{1} \, e^{V^{(0)}_{1}+V^{(0)}_{2}}\Phi^{(0)}_1\,\delta c_{g[0]}=0\,,\label{eq:horizon_vector_contraint_o2}
\end{align}
while solving the equation of motion \eqref{eq:bulk_eom_chi} at the same order yields the asymptotic expansion
\begin{align}
\delta\chi_{1[1]}&=\frac{r^{2\Delta_{Z_{1}} -3}}{(2\Delta_{Z_{1}}-3)\phi_{1v}^{2}}\,\left\{-e^{V^{(0)}_{1}+V^{(0)}_{2}} \Phi^{(0)}_1\, \left(i\omega_{[2]} \delta c_{g[0]} - k_{1}v_{[2]}\right)+q^2\,\delta c_{g[0]} \int dr\,e^{V_2-V_1}\Phi_1 \right.\nn
&\left. + iq\,k_{1} \int dr\,e^{V_2-V_1}\left(\partial_{\phi_I} \Phi_1 \, \delta\phi_{I[1]} + \partial_{\psi_I} \Phi_1 \, \delta\psi_{I[1]} +\Phi_1 \left(\delta V_{2[1]}-\delta V_{1[1]}\right)\right) \right\}+\cdots\,.\notag
\end{align}
Comparing once again with our general asymptotics \eqref{eq:gen_UVexp}, the $\zeta_{S_{1}}=0$ condition gives
\begin{align}
&e^{V^{(0)}_{1}+V^{(0)}_{2}} \Phi^{(0)}_1\, \left(i\omega_{[2]} \delta c_{g[0]} - k_{1}v_{[2]}\right) -q^2\,\delta c_{g[0]} \int dr\,e^{V_2-V_1}\Phi_1\nn
&+ iq\,k_{1} \int dr\,e^{V_2-V_1}\left(\partial_{\phi_I} \Phi_1 \, \delta\phi_{I[1]} + \partial_{\psi_I} \Phi_1 \, \delta\psi_{I[1]} +\Phi_1 \left(\delta V_{2[1]}-\delta V_{1[1]}\right)\right) =0\,.\label{eq:bulk_eom_chi1o2_int}
\end{align}
Like at the previous order in $\varepsilon$, we proceed to eliminate the horizon fluid velocity $v_{[2]}$ by using equation \eqref{eq:horizon_vector_contraint_o2} to obtain
\begin{align}\label{eq:omega_2eq}
&\left(e^{V^{(0)}_{1}+V^{(0)}_{2}} \Phi^{(0)}_1\,\frac{k_{s1}^2\Psi^{(0)}_1}{\mathcal{B}}\omega_{[2]} +iq^2\,\frac{\delta^2 w_{FE}}{\delta k_{1}^2}\right)\delta c_{g[0]}\nn
& +q\left(k_{1}\Phi^{(0)}_1\frac{s}{\mathcal{B}} -\frac{\delta s}{\delta k_{1}}\right)\delta T_{[1]} +q\left(k_{1}\Phi^{(0)}_1\frac{\rho}{\mathcal{B}} -\frac{\delta \rho}{\delta k_{1}}\right)\delta \mu_{[1]}=0\,.
\end{align}
In addition to this, we make use of the scalar constraint equations \eqref{hconstrainta}, \eqref{hconstrainttwo} at third order,
\begin{align}\label{thorderce2}
&i\omega_{[2]}(T^{-1}c_\mu \delta T_{[1]} +\xi \delta \mu_{[1]}) -q^2 \frac{4\pi \left(s\delta T_{[1]} + \rho\delta\mu_{[1]}\right)}{\mathcal{B}} +\omega_{[2]}q\left(\frac{k_{1}\Phi^{(0)}_1s}{\mathcal{B}} -\frac{\delta s}{\delta k_{1}}\right)\delta c_{g[0]} =0\,,\nn
&-q^2e^{V^{(0)}_{2}-V^{(0)}_{1}}\tau^{(0)}\delta\mu_{[1]} -q^2\rho \, e^{-V^{(0)}_{1}-V^{(0)}_{2}}\frac{\left(s\delta T_{[1]} + \rho\delta\mu_{[1]}\right)}{\mathcal{B}} +i\omega_{[2]} \left(\xi \delta T_{[1]} +\chi \delta \mu_{[1]}\right) \nn
& +\omega_{[2]}q \left(\frac{k_{1}\Phi^{(0)}_1\rho}{\mathcal{B}} -\frac{\delta\rho}{\delta k_{1}}\right)\delta c_{g[0]}=0\,,
\end{align}
where we have already substituted for $v_{[2]}$. The equations \eqref{eq:omega_2eq} and \eqref{thorderce2} form the $3\times3$ system in the main text, \eqref{3by3}, which can be solved for three solutions of $\omega_{[2]}$ to give the diffusion constants of the three modes.

\section{Heat current}\label{app:heatcurrent}
In this appendix we construct the bulk and boundary heat currents following \cite{Banks:2015wha}.  Let us consider a general vector $k^\mu$. We define the $2$-form
\begin{align}\label{eq:g2form}
G^{\mu\nu}= -2\nabla^{[\mu} k^{\nu]} - \tau\, k^{[\mu} F^{\nu]\rho} A_\rho -\frac{1}{2}\left(k^\rho A_\rho-f\right) \tau F^{\mu\nu}\,,
\end{align}
where $k^\mu F_{\mu\nu} = \partial_\nu f +\beta_\nu$, with $\beta$ a $1$-form and $f$ a globally defined function. Using the identity
\begin{align}\label{eq:kidentity}
\nabla_{\mu}\nabla^{[\mu}k^{\nu]} = -R^{\nu}{}_{\rho} k^{\rho} -\nabla^{\nu}\nabla_{\rho}k^{\rho} +\nabla_{\mu}\nabla^{(\mu}k^{\nu)}\,,
\end{align}
the equations of motion \eqref{eq:eom1} imply that
\begin{align}\label{eq:nablag2form}
\nabla_\mu G^{\mu\nu}&=V\, k^{\nu} +2\nabla^{\nu}\nabla_{\rho}k^{\rho} -2\nabla_{\mu}\nabla^{(\mu}k^{\nu)} +\frac{1}{2}\tau F^{\nu\rho}\beta_\rho -\frac{1}{2}A_\rho \mathcal{L}_k \left(\tau\,F^{\nu\rho}\right)-\frac{\tau}{2}\,F^{\nu\rho}A_{\rho}\,\nabla_{\mu}k^{\mu}\nn
&+\left(\sum_{I}\Phi_{I}\,\partial^\nu\chi^{I}\partial_\rho\chi^{I} +\sum_{J}\Psi_{J}\,\partial^\nu\sigma^{J}\partial_\rho\sigma^{J}\right) k^\rho\nn
&+\left(\sum_{I} G_{I}\,\partial^\nu\phi^{I}\partial_\rho\phi^{I} +\sum_{J} W_{J}\,\partial^\nu\psi^{J}\partial_\rho\psi^{J}\right) k^\rho \,.
\end{align}
%where the dots indicate corrections of order $\mathcal{O}(\varepsilon^3)$. 
Let us consider $k^\mu=\partial_t$ and a general static background metric of the form
\begin{align}\label{eq:appB_metric}
ds^{2}=-G\,dt^{2}+F\,dr^{2}+\tilde{g}_{ij}\,dx^{i}dx^{j}\,,
\end{align}
with the functions $G$, $F$ and the $d-1$ dimensional metric $\tilde{g}_{ij}$ depending an all coordinates except for $t$. The bulk heat current is defined as
\begin{align}\label{eq:heat_def_app}
\delta Q^i_{bulk} = \sqrt{-g}G^{ir} &=\frac{G^{3/2}}{F^{1/2}}\sqrt{\tilde{g}}\,\tilde{g}^{ij}_d\left(\partial_r\left(\frac{\delta g_{jt}}{G}\right)-\partial_j\left(\frac{\delta g_{rt}}{G}\right)
\right) -a_t\, \delta J_{bulk}^i\,,\nn
&=G^{1/2} \sqrt{\tilde{g}} \left[2 K^i{}_t+F^{-1/2} \tilde{g}^{ij}\,\partial_t \delta g_{rj}(t,r,x^i)\right] -a_t \,\delta J_{bulk}^i\,,
\end{align}
where we have used the result of Appendix B of \cite{Donos:2017gej} for the extrinsic curvature component
\begin{align}\label{eq:extr_curv}
K^i{}_{\,t}&=\frac{1}{2} G\,F^{-1/2} \tilde{g}^{ij} \left[ \partial_r \left(\frac{\delta g_{jt}}{G}\right) - \partial_j \left(\frac{\delta g_{rt}}{G}\right)-\frac{\partial_t \delta g_{rj}}{G}\right]\,.
\end{align}
Note that if we write $\tilde{t}^\mu{}_\nu=-2 K^\mu{}_\nu+X \delta^\mu{}_\nu+Y^\mu{}_\nu$, where $X=2K+\cdots$ and $Y$ are additonal terms that come from the counterterms, then $\tilde{t}^\mu{}_\nu$ gives the stress tensor when evaluated on the boundary. Thus

\begin{align}
\delta Q^i_{bulk} =G^{1/2} \sqrt{\tilde{g}}\left [Y^i{}_t- \tilde{t}^i{}_t+F^{-1/2} \tilde{g}^{ij}\ \partial_t \delta g_{rj}(t,r,x^i)\right]-a_t \,\delta J_{bulk}^i\,.
\end{align}

Evaluated on the boundary, this gives

\begin{align}
\delta Q_{bulk}^i\Big|_\infty =-  \left(r^{-2} \,t^i{}_t+\mu\, \delta J_{bulk}^i\Big|_\infty\right)\,,
\end{align}
where $t^\mu{}_\nu=r^5 \tilde{t}^\mu{}_\nu$. Note that the contribution from $Y^i{}_t$, as coming from \eqref{eq:bdy_action}, and contribution from the term involving a time derivative are subleading even in the precense of sources. This result matches the expression for the boundary heat current obtained from the variation of the action in the presence of the sources \eqref{eq:ezeta_sources}
\begin{align}
\delta S&=\int d^3x  \,\sqrt{-h}\,\left[ \frac{1}{2}\,r^{-5}\,t^{\mu\nu}\,\delta g_{\mu\nu}+r^{-3} J^\mu\delta A_\mu\right]\,,
\end{align}
where $h_{\mu\nu}=g_{\mu\nu}-n_\mu\,n_\nu$ and $n$ is the unit norm normal vector. Furthermore, equation \eqref{eq:nablag2form} implies the radial dependence

\begin{align}\label{eq:drheat_bulk}
\partial_r \delta Q_{bulk}^i & = \partial_j \left(\sqrt{-g}G^{ji}\right) + \partial_t \left(\sqrt{-g}G^{ti}\right) -\frac{1}{2}\sqrt{-g}\,\tau F^{i\rho}\partial_t A_\rho +\frac{1}{2}\sqrt{-g}\,A_\rho \partial_t \left(\tau\,F^{i\rho}\right) \nn
& -2\sqrt{-g}\,\partial^i\partial_t\log\sqrt{-g} +\partial_{\mu}\left(\sqrt{-g}\nabla^{i}k^{\mu}\right) -\sqrt{-g}\left(\sum_{I}\Phi_{I}\,\partial^i\chi^{I}\partial_t \chi^{I} +\sum_{J}\Psi_{J}\,\partial^i \sigma^{J}\partial_t\sigma^{J}\right)\nn
&-\sqrt{-g} \left(\sum_{I} G_{I}\,\partial^i \phi^{I}\partial_t \phi^{I} +\sum_{J} W_{J}\,\partial^i \psi^{J}\partial_t \psi^{J}\right)+\frac{\tau}{2}F^{i\rho}A_{\rho}\,\partial_{t}\sqrt{-g}\,.
\end{align}
One can check that for our choice of $k^\mu$, only the last term of the second line in \eqref{eq:drheat_bulk} contributes to order $\mathcal{O}\left(\varepsilon^2\right)$, which leads to the radial evolution for $\delta Q^1_{bulk}$ presented in \eqref{eq:drheat_bulk_1}, and the relation \eqref{eq:bdy_hor_heat} between the boundary and horizon heat currents.

\section{$\varepsilon$-expansion in the presence of sources}\label{app:hydro}
The aim of this appendix is to presence some of the calculations that were omitted in section \ref{sec:hydro}. In particular, as it was explained in the main text, we will repeat the $\varepsilon$-expansion of section \ref{sec:pert} but in the presence of the external sources \eqref{eq:ezeta_sources}. In these computations we keep $\omega_{[1]}=\delta T_{[0]}=\delta \mu_{[0]}=0$ following the results of Appendix \ref{app:hydro_expansion}.

\subsection{Vector constraint and equation of motion for $\chi_1$}

We start with the $\varepsilon$-expansion of the vector constraint. Evaluating \eqref{hiconstraint2b} at order $\mathcal{O}(\varepsilon^2)$ we obtain
\begin{align}
& e^{V_1^{(0)} + V_2^{(0)}} \mathcal{B} v_{[2]} + iq \left(s\delta T_{[1]} + \rho\delta\mu_{[1]}\right) -i\omega_{[2]} \,k_{1} e^{V_1^{(0)} + V_2^{(0)}}\Phi_1^{(0)}\,\delta c_{g}- sT\,\zeta-\rho\, E=0\,. \label{eq:vc_sourced}	
\end{align} 
The above equation can used to determine $v_{[2]} $ in terms of the sources, $\delta c_g$ and thermodynamic quantities.

We now move on to discuss the equation of motion for $\chi_1$ \eqref{eq:bulk_eom_chi}. The first non-trivial terms appear at order $\mathcal{O}(\varepsilon^2)$ giving
 \begin{align}\label{eq:bulk_eom_chi1o2}
\partial_{r}\left[e^{V_1 + V_2} \, \Phi_1\, \left(-i\omega_{[2]} \delta c_{g} + U\partial_{r}\delta\chi_{1[2]} - k_{1}e^{-2 V_1}\, \delta g_{r1[2]}\right)\right]+ e^{V_2 - V_1}\Bigl[ -q^2\,\Phi_1\, \delta c_{g} &\nn
+ iq \, \partial_{\phi_I} \Phi_1 \phi_{I[1]} + iq \, \partial_{\psi_I} \Phi_1 \, \psi_{I[1]} + iq \, k_{1} \,\frac{\Phi_1}{2} \, \left(- e^{-2V_{1}}\delta g_{11[1]} + e^{-2V_{2}}\delta g_{22[1]} \right) -k_1 \Phi_1 \zeta\Bigr]= 0&\,.
\end{align}
The asymptotic behaviour of the solution for $\delta\chi_{1[2]}$ is
\begin{align}\label{eq:chi12_as_exp}
\delta \chi_{1[2]} &=\frac{r^{2\Delta_{Z_1}-3}}{(2\Delta_{Z_1}-3)\phi_{1v}^2}\Bigl[ e^{V_1^{(0)} + V_2^{(0)}}\Phi_1^{(0)} \left(-i \omega_{[2]}\delta c_g+k_1 v_{[2]}\right) +\zeta \,w^1\nn
&+iq \left(\nu^1\, \delta T_{[1]} +\beta^1\, \delta \mu_{[1]} -iq\,w^{11}\, \delta c_{g[0]} \right) \Bigr] +\cdots\,,
\end{align}
where we have used the definition of the susceptibilities. Imposing that $\zeta_{S_{1}}=0$ gives the equation
\begin{align}
&e^{V_1^{(0)} + V_2^{(0)}} \Phi_1^{(0)}\, \left(i\omega_{[2]} \delta c_{g} - k_{1} v_{[2]}\right) - iq \left(\nu^1 \delta T_{[1]}+ \delta \mu_{[1]}\,\beta^1-i\,q\,\delta c_{g}\, w^{11} \right) -\zeta\, w^1 = 0\,. \label{eq:chieom_sourced}
\end{align}

\subsection{Constitutive relation for boundary $U(1)$ current}
It is straightforward to show that the bulk $U(1)$ current, defined as $\delta J^i_{bulk}
\equiv\sqrt{-g} \tau \delta F^{ir}$, is given by 
\begin{align}
\delta J^i_{bulk}&= \varepsilon^2 \,e^{V_2-V_1}\, \tau\, e^{-i\omega_{[2]} v_{EF} + iq x^1} \left[E-a\zeta -U\partial_{r}\delta a_{1[2]} - iq\,U\, \partial_r g\,\delta\mu_{[1]} - \delta g_{t1[2]} \partial_{r}a \right] +\mathcal{O}(\varepsilon^3)\,,
\end{align}
where the function $g(r)$ was defined above \eqref{eq:nh_static_metric_pertb}. Furthermore, using Maxwell's equations \eqref{eq:eom1}, it can be shown that
\begin{align}\label{eq:bulk_j_eom}
\partial_r \delta J^i_{bulk}=\partial_j \left(\sqrt{-g} \tau \delta F^{ji}\right) + \partial_t \left(\sqrt{-g} \tau \delta F^{ti}\right)=0+\mathcal{O}(\varepsilon^3) \,.
\end{align}
Thus, the boundary current is related to the horizon current by
\begin{equation}
\delta J^1=\lim_{r\to\infty} \delta J^1_{bulk}=e^{-i\omega_{[2]} v_{EF} + iq x^1}\, J^{1}_{(0)}+\mathcal{O}(\varepsilon^3)\,.
\end{equation}
This gives the constitutive relation \eqref{eq:J_constitutive} when we eliminate $v_{[2]}$ using the constraint \eqref{eq:vc_sourced}.

\subsection{Constitutive relation for boundary heat current}
From the definition of the bulk heat current \eqref{eq:heat_def_app} in appendix \ref{app:heatcurrent}, it is easy to show that
\begin{align}\label{eq:heat_def}
\delta Q^1_{bulk} &= \varepsilon^2 e^{V_2-V_1} e^{-i\omega_{[2]} v_{EF} + iq x^1} \left(  -\zeta\,U + U\partial_r \delta g_{t1[2]} - \delta g_{t1[2]}\,\partial_r U - iq U\,\delta g_{rt[1]} \right) -a \delta J^i_{bulk} +\mathcal{O}(\varepsilon^3)\,,
\end{align}
where we have used that for our background \eqref{eq:DC_ansatz} we need to set $G=F^{-1}=U$ and $\tilde{g}_{ij}=e^{2V_{i}}\delta_{ij}$ in \eqref{eq:appB_metric}. Following from \eqref{eq:heat_def_app}, this bulk quantity satisfies
\begin{align}\label{eq:drheat_bulk_1}
\partial_r \delta Q^1_{bulk} & = \varepsilon^2 e^{V_2-V_1}\,e^{-i\omega_{[2]} v_{EF} + iq x^1}\, \Phi_1\,k_1\,i\omega_{[2]} \delta c_g +\mathcal{O}(\varepsilon^3)
\end{align}
and thus the boundary heat current is related to the horizon heat current by
\begin{align}\label{eq:bdy_hor_heat}
\delta Q^1 =\lim_{r\to\infty} \delta Q^1_{bulk}&=e^{-i\omega_{[2]} v_{EF} + iq x^1} \left(Q^1_{(0)} +i\omega_{[2]}
\delta c_g \,w^1\right)+\mathcal{O}(\varepsilon^3)\nn
& = e^{-i\omega_{[2]} v_{EF} + iq x^1} \left(sT\, v_{[2]} +i\omega_{[2]}\,\delta c_g \,w^1\right)+\mathcal{O}(\varepsilon^3)\,. 
\end{align}
 As before, we can use the constraint \eqref{eq:vc_sourced} to eliminate $v_{[2]}$ from the expression above, leading to the constitutive relation \refeq{eq:Q_constitutive}.

\section{Weak momentum relaxation limit}\label{app:weak_mom_relax}
In this appendix we will study the source-free dynamics of the theory we introduced in sections \ref{sec:hydro} and \ref{sec:pinning}. We start by eliminating the time derivatives of $\delta\hat{c}_{g}$ from the constitutive relations for the currents \eqref{eq:J_constitutive} and \eqref{eq:Q_constitutive} by using the Josephson relation \eqref{eq:dy_source} to obtain the expressions,
\begin{align}
\delta Q^{1}=&-\left( T\,\bar{\alpha}_{H}+\frac{\upsilon}{\gamma}\,\Delta\,\rho\,T\,s-\upsilon\,\Delta\,s\,T\,k_{1}\,\beta^{1}\right)\partial\delta \hat{\mu} -\left( \bar{\kappa}_{H}+\frac{\upsilon}{\gamma}\,\Delta\,T\,s^{2}-\upsilon\,\Delta\,s\,T\,k_{1}\,\nu^{1}\right)\partial\delta \hat{T}\notag\\
&-\upsilon\,w^{11}\,k^{2}_{1}\,\Delta\,T s\left(-\partial^{2}+l^{2}\right)\,\delta\hat{\phi}\,,\label{eq:Q_constitutiveV2}\\
\delta J^{1}=&-\left( \sigma_{H}+\frac{1}{\gamma}\,\Delta\,\rho^{2}-\Delta\,\rho\,k_{1}\,\beta^{1}\right)\partial \delta\hat{\mu}-\left( \alpha_{H}+\frac{1}{\gamma}\,\Delta\,\rho\,s-\Delta\,\rho\,k_{1}\,\nu^{1}\right)\partial \delta \hat{T}\notag\\
&-w^{11}\,k^{2}_{1}\,\Delta\,\rho\left(-\partial^{2}+l^{2}\right)\,\delta\hat{\phi}\label{eq:J_constitutiveV2}\,.
\end{align}
In the expressions above we have defined
\begin{align}
&\Delta=\frac{4\pi}{s\,k_{s1}^{2}\,\Psi_{1}^{(0)}},\qquad l^{2}=\frac{\left| \mathcal{O}_{Z}\right|\,\delta\phi_{1s}}{w^{11}},\qquad \upsilon=1+\frac{\mathcal{B}\,w^{1}}{k_{1}\Phi_{1}^{(0)}Ts}\,\notag\\
&\gamma=1+\frac{k_{s1}^{2}\,\Psi_{1}^{(0)}}{k_{1}^{2}\,\Phi_{1}^{(0)}},\qquad \delta\hat{c}_{g}=-k_{1}\,\delta\hat{\phi}\,.
\end{align}
The Josephson relation \eqref{eq:dy_source} now reads
\begin{align}\label{eq:Josephson}
\partial_{t}\delta\hat{\phi}=&-\Delta \left(\gamma k_{1}^{2} w^{11} (-\partial^{2}+l^{2} )\,\delta\hat{\phi}+(\rho-k_{1}\gamma \beta^{1})\,\partial\delta\hat{\mu}+(s-k_{1} \gamma \nu^{1})\,\partial\delta\hat{T}\right)\,.
\end{align}
The advantage of this parametrisation is that at weak momentum relaxation we will have $\gamma\to1$ and $\Delta$ parametrically large. At this point we stress that even when we take a weak momentum relaxation limit, our theory is valid for frequencies and pinning strength much smaller than the scale $1/\Delta$. In this limit we have
\begin{align}
\delta Q^{1}=&-\Delta\,\upsilon\left(\rho\,T\,s-s\,T\,k_{1}\,\beta^{1}\right)\partial\delta \hat{\mu} -\Delta\,\upsilon \left( T\,s^{2}-s\,T\,k_{1}\,\nu^{1}\right)\partial\delta \hat{T}\notag\\
&-\upsilon\,\Delta\,w^{11}\,k^{2}_{1}\,T s\left(-\partial^{2}+l^{2}\right) \delta\hat{\phi}\,,\label{eq:Q_constitutiveV2weak}\\
\delta J^{1}=&-\Delta \left( \rho^{2}-\rho\,k_{1}\,\beta^{1}\right)\partial \delta\hat{\mu}-\Delta\left( \rho\,s-\rho\,k_{1}\,\nu^{1}\right)\partial \delta \hat{T}\notag\\
&-\Delta\,w^{11}\,k^{2}_{1}\,\rho\left(-\partial^{2}+l^{2}\right)\delta\hat{\phi}\label{eq:J_constitutiveV2weak}\,,
\end{align}
while all the terms on the right hand side of \eqref{eq:Josephson} are leading order in $\Delta$.

We will now obtain a theory of incoherent hydrodynamics starting from the description of \cite{Delacretaz:2017zxd} which concerns systems with small phase and momentum relaxations rates $\kappa_{n}k_{0}^{2}\xi$ and $\Gamma$ respectively. As we will see, the main disagreement will arise from issues related to  thermodynamics. More specifically, the systems considered there have $w^{1}=\beta^{1}=\nu^{1}=0$ which is  not true in general for holographic theories. This fact has an effect even in the limit of weak momentum relaxation as can be seen from equations \eqref{eq:Q_constitutiveV2weak}, \eqref{eq:J_constitutiveV2weak} and \eqref{eq:Josephson}.

In the limit $\omega\sim\kappa_{n}k_{0}^{2}\xi\ll \Gamma$ we can use the momentum conservation equation to express the fluid velocity locally as
\begin{align}
v=-\frac{1}{\Gamma\chi_{\pi\pi}}\,\left( \rho\,\partial\delta\mu+s\,\partial\delta T+\kappa_{n}\,(-\partial^{2}+k_{0}^{2})\,\delta\phi\right)\,.
\end{align}
Plugging this expression in their constitutive relations for the electric and heat current we find,
\begin{align}
\delta Q^{1}&=-\left(\kappa_{0}+\frac{T\,s^{2}}{\Gamma\chi_{\pi\pi}} \right)\partial \delta T-T\left( \alpha_{0}+\frac{s\rho}{\Gamma\chi_{\pi\pi}}\right)\partial \delta\mu+T\kappa_{n}\left( \gamma_{2}-\frac{s}{\Gamma\chi_{\pi\pi}}\right)\,\left(-\partial^{2}+k_{0}^{2}\right)\delta\phi\,, \label{eq:Q_constitutiveHB}\\
\delta J^{1}&=-\left(\sigma_{0}+\frac{\rho^{2}}{\Gamma\chi_{\pi\pi}} \right)\partial \delta \mu-\left( \alpha_{0}+\frac{s\rho}{\Gamma\chi_{\pi\pi}}\right)\partial \delta T+\kappa_{n}\left( \gamma_{1}-\frac{\rho}{\Gamma\chi_{\pi\pi}}\right)\,\left(-\partial^{2}+k_{0}^{2}\right)\delta\phi\label{eq:J_constitutiveHB}\,,
\end{align} 
while for the Josephson relation we have
\begin{align}\label{eq:JosephsonHB}
\partial_{t}\delta\phi=\left( \gamma_{1}-\frac{\rho}{\Gamma\chi_{\pi\pi}}\right)\partial\delta\mu+\left( \gamma_{2}-\frac{s}{\Gamma\chi_{\pi\pi}}\right)\partial\delta T-\kappa_{n}\left( \xi+\frac{1}{\Gamma\chi_{\pi\pi}}\right)\left(-\partial^{2}+k_{0}^{2}\right)\delta\phi\,.
\end{align}

We see that the ratio of the coefficient of $\delta\phi$ in \eqref{eq:Q_constitutiveHB} and the coefficient of $\partial\delta T$ in \eqref{eq:JosephsonHB} is the same with ratio of the coefficient of $\delta\phi$ in \eqref{eq:J_constitutiveHB} and the coefficient of $\partial\delta \mu$ in \eqref{eq:JosephsonHB} times $T$. This is true even without taking a limit of small momentum relaxation rate $\Gamma$. In general, this constraint is not satisfied in the system \eqref{eq:Q_constitutiveV2weak}, \eqref{eq:J_constitutiveV2weak} and \eqref{eq:Josephson}. As we mentioned earlier, the discrepancy comes from thermodynamic factors of $\beta^{1}$, $\nu^{1}$ and $w^{1}$ which are non-zero for our holographic theory.

The fact that $w^{1}$ is non-zero for our model is related to the fact that we are examining branches of black holes which are not thermodynamically preferred. Indeed, the thermodynamically dominant configurations will minimise the free energy of the system and therefore they must have $w^{1}=0$. For the specific model we are considering, the preferred branch will also have $\beta^{1}=\nu^{1}=0$ but this point is not true for more general holographic theories which exhibit spontaneous breaking of translations, see e.g. \cite{Donos:2012wi,Withers:2013kva,Donos:2013woa,Amoretti:2017frz}. For such theories one might expect new terms to arise in the constitutive relations but the terms we have identified in sections \ref{sec:hydro} and \ref{sec:pinning} for our minimal model \eqref{eq:bulk_action} will still be present.

%\newpage
\bibliographystyle{utphys}
%\bibliography{refs}{}
\providecommand{\href}[2]{#2}\begingroup\raggedright\endgroup

\end{document}